\documentclass[preprint,largeabstract]{aastex}

\shorttitle{Aerosol properties in EGPs}
\shortauthors{Lavvas $\&$ Koskinen}

 \usepackage{lineno}
 \usepackage{color}	
 \newcommand{\clr}{black}
 
\begin{document}


\title{Aerosol properties in the atmospheres of extrasolar giant planets}


\author{P. Lavvas}
\affil{{\color{\clr} Groupe de Spectrom\'etrie Mol\'eculaire et Atmosph\'erique, UMR CNRS 7331,} \\ Universit\'e de Reims Campagne Ardenne, Reims, France}
\email{panayotis.lavvas@univ-reims.fr}

\and

\author{T. Koskinen}
\affil{Lunar and Planetary Laboratory, University of Arizona, Tucson, USA}

%
%\author{R. J. Hanisch\altaffilmark{5}}
%\affil{Space Telescope Science Institute, Baltimore, MD 21218}


%\altaffiltext{1}{Visiting Astronomer, Cerro Tololo Inter-American Observatory.
%CTIO is operated by AURA, Inc.\ under contract to the National Science
%\altaffiltext{2}{Society of Fellows, Harvard University.}
%\altaffiltext{3}{present address: Center for Astrophysics,
%    60 Garden Street, Cambridge, MA 02138}
%\altaffiltext{4}{Visiting Programmer, Space Telescope Science Institute}
%\altaffiltext{5}{Patron, Alonso's Bar and Grill}


\begin{abstract}

{\color{\clr}We use a model of aerosol microphysics to investigate the impact of high-altitude photochemical aerosols on the transmission spectra and atmospheric properties of close-in exoplanets, such as HD209458b and HD189733b. The results depend strongly on the temperature profiles in the middle and upper atmosphere that are poorly understood. Nevertheless, our model of HD189733b, based on the most recently inferred temperature profiles, produces an aerosol distribution that matches the observed transmission spectrum. We argue that the hotter temperature of HD209458b inhibits the production of high-altitude aerosols and leads to the appearance of a more clear atmosphere than on HD189733b. The aerosol distribution also depends on the particle composition, the photochemical production, and the atmospheric mixing. Due to degeneracies among these inputs, current data cannot constrain the aerosol properties in detail. Instead, our work highlights the role of different factors in controlling the aerosol distribution that will prove useful in understanding different observations, including those from future missions. For the atmospheric mixing efficiency suggested by general circulation models (GCMs) we find that aerosol particles are small ($\sim$nm) and probably spherical. We further conclude that composition based on complex hydrocarbons (soots) is the most likely candidate to survive the high temperatures in hot Jupiter atmospheres. Such particles would have a significant impact on the energy balance of HD189733b's atmosphere and should be incorporated in future studies of atmospheric structure. We also evaluate the contribution of external sources in the photochemical aerosol formation and find that their spectral signature is not consistent with observations.}

\end{abstract}





%\linenumbers

\section{Introduction}

%general intro
The study of exoplanetary atmospheres is rapidly expanding providing new insights to the complexity and diversity of the atmospheres of these distant planets. Apart from the chemical inventories observed on these objects, recent studies suggest that subsequent products of chemical evolution such as hazes can also exist in these atmospheres. Hazes can strongly influence the thermal structure, the dynamics, and the photochemistry, as observations of solar system atmospheres have demonstrated \citep{West09,West13}. {\color{\clr}Depending on their formation mechanism hazes can be separated into photochemical or condensate nature. Here we investigate the possible properties of hazes of photochemical origin} in the atmospheres of extrasolar giant planets, based on available observations. 

%observations of aerosols
Among the most observed exoplanets are the hot-Jupiters HD 209458 b and HD 189733 b.
Primary eclipse observations of HD 189733 b at UV, visible and IR wavelengths by different instruments of the Hubble Space Telescope (HST) and Spitzer suggest a rather monotonic decrease in the transit depth with increasing wavelength \citep{Pont08,Sing09,Sing11,Sing16}. This wavelength dependence is contrary to the sharp molecular signatures anticipated at near-IR wavelengths by H$_2$O, CH$_4$, CO and CO$_2$, which appear in secondary eclipse observations of this atmosphere \citep{Grillmair08,Desert09,Swain09}. Primary transit observations probe higher altitudes than secondary eclipse, therefore these observations were interpreted as an indication for the presence of a silicate condensate (Enstantite, MgSiO$_3$) at the pressure levels probed \citep{Lecavelier08}. However, subsequent studies of cloud microphysics demonstrate that silicate clouds can not easily reproduce the observed primary transit signature \citep{Lee15, Pinhas17}, thus a different type of heterogeneous opacity is required.

On the other hand, the primary transit spectrum of HD 209458 b includes signatures of Rayleigh scattering, Na and H$_2$O and it is roughly consistent with a clear atmosphere model \citep{Lavvas14}. HD 209458 b is known to have a very low visible albedo \citep{Rowe06} and this characteristic could be caused by absorption from TiO or VO \citep{Hubeny03,Desert08,Lecavelier08b}, which are known to have large absorption cross sections at visible wavelengths, and can be present in the deep troposphere. {\color{\clr}Similarly, any gaseous species absorbing in the visible with an absorption opacity larger than the scattering opacity by clouds would have the same effect \citep{Marley99, Sudarsky00}.} Theoretical temperature profiles including heating by such molecules \citep{Showman09} are consistent with the secondary eclipse observations \citep{Diamond14,Line16a}. Other possible visible extinction candidates include sulfur photochemical products \citep{Zahnle09}, and silicate clouds \citep{Fortney03}. Therefore, current observations and modelling suggest that a high-altitude heterogeneous opacity similar to that on HD 189733 b is not required to explain observations of HD 209458 b.

It is clear that the analysis of these observations is difficult due to the presence of  large systematic and random uncertainties for the different instruments involved. For example, the near-IR observations of HD 189733 b do suggest the presence of some gaseous absorbers, but the uncertainties of the analysis do not provide a clear picture \citep{Swain08,Gibson11}. Therefore a definite interpretation of the available observations might prove equivocal, and repeated observations are required to resolve such issues. Notwithstanding these obstacles we can investigate the properties of hazes in hot-Jupiter atmospheres and potentially aid in the interpretation of the observations. 

%goal of study
Given the limited information we have for hazes in exoplanet atmospheres, we can first consider what we know about hazes from the atmospheres of the solar system. Numerous studies have demonstrated the multiple ways in which hazes can affect atmospheric evolution and planetary surface properties, from short to long term aspects. The most characteristic example is our own planet (through climate change), as well as, Saturn's moon Titan, where hazes are the dominant feature of the atmosphere and are known to affect the thermal structure, the photochemistry, and the surface properties \citep[see recent review by][]{West13}. The most relevant examples though are the cases of the giant planets, which are also affected by different kinds of hazes. 

{\color{\clr} On Uranus and Neptune, scattering by small haze particles affects the albedo and the thermal structure of their atmospheres \citep{Marley99b}. On Jupiter and Saturn for which more observations are available}, hazes can be separated into two general families: photochemical aerosols and clouds \citep[][]{West04,West09}. Photochemical aerosols are present in the stratosphere and upper troposphere of both giant planets, while clouds are formed at deeper regions. The composition of photochemical aerosols has not been clearly identified yet. Both observations and modelling suggest that they are hydrocarbon based \citep{Jaffel95, Wong03,Koskinen16}, but contributions from phosphine chemistry on Saturn and sulfur chemistry on Jupiter are also anticipated, although not yet verified. Average particle sizes range between $\sim$0.01 $\mu$m in the stratosphere to $\sim$0.1 $\mu$m in the upper troposphere. At deeper layers the condensation of ammonia, ammonia sulfide, and water dominate the particulate extinction in these atmospheres, and photochemical aerosols produced at higher altitudes act as nucleation sites for the production of cloud particles that eventually grow to sizes of a few microns. A similar stratification can be anticipated for hazes in exoplanet atmospheres, although the chemical composition of the aerosols and the gaseous species condensing into clouds could be different \citep{Marley13}. 

Given our current understanding for the role of hazes in planetary atmospheres, and motivated by the above observations of HD 189733 b, we would like to understand how hazes would behave in exoplanet atmospheres. In principle, such an investigation would require detailed studies addressing multiple aspects of haze properties, which can be specific for each planet. However, at the current early stage of this investigation we can first attempt to address basic questions common to all exoplanet cases, such as:
\begin{enumerate}
\item[-]  What kind of hazes forms in exoplanet atmospheres and under which conditions? Why do they seem to be present in one case (HD 189733 b) and absent in another (HD 209458 b)?
\item[-] What are the processes that define the production and evolution of hazes in these environments, and how do these compare with what we already know from the atmospheres of our solar system?
\item[-] If hazes do exist in exoplanet atmospheres, what would be their properties regarding their size, their density and their optical properties? What methods are required to detect them in future observations? 
\item[-] How would the hazes interact with their atmospheric background? What would be their role in the thermal structure of the atmosphere, their interaction with atmospheric circulation, and how would they affect the atmospheric composition (e.g. heterogeneous chemistry, photochemistry)?
\end{enumerate}

Definite answers to these question are restrained by the limited number of observations currently available, as well as the multitude environmental conditions among various planets. Nevertheless, we can utilise the methods and knowledge we have gained from solar system studies, along with the current understanding of exoplanet atmospheric properties, in order to address some of the above questions. 

{\color{\clr} We note that \cite{Helling06} have also presented detailed models for the formation and evolution of clouds in drown dwarf atmospheres. These were subsequently applied to the atmospheric conditions of EGPs in order to investigate the composition and optical properties of clouds in these environments in relation to available observations \citep{Lee15}, the impact of the clouds on the atmospheric structure \citep{Lee17}, and the possible occurrence of lightning in these cloud structures \citep{Helling13}. Instead of the physics of cloud formation, in the current study we focus on the properties} of the photochemical aerosols that can form in the upper atmosphere {\color{\clr}of EGPs}. 

We use a 1D aerosol microphysics model that we adapted to the environments of EGPs. We apply this model to the two representative cases of EGPs: HD 209458 b and HD 189733 b, for which we have a better understanding of atmospheric properties relative to other planets. We note that our study focuses only on the photochemically produced aerosols and does not follow the interaction of these aerosols with the background gases, which can condense and lead to the formation of cloud particles. As outlined by the nature of the questions raised above, our goal in this study is not to provide a detailed analysis for each planet separately, but more interestingly to investigate the phase space of resulting aerosols properties in such atmospheres. This investigation provides a basic background for aerosol properties in exoplanet atmospheres, on which planet-specific investigations will further advance our knowledge. 

\section{Model description}

We use a 1D aerosol microphysics model to investigate the properties of aerosol particles in extrasolar atmospheres. Our prototype model is described in detail in \cite{Lavvas10}, where it was used for the interpretation of Titan's aerosol $in~situ$ observations obtained with the Huygens probe. Here the model is updated to represent the atmospheric conditions found on EGPs. In a nutshell, the model utilises a geometrically expanding grid of particle sizes and calculates the population of each particle size by solving the continuity equation:
\begin{eqnarray}
{ \frac{\partial{n_p }} {\partial{t}} = -\frac{1}{r^2} \frac{\partial{(r^2\Phi(v_p))}}{\partial{r}} } \nonumber \\
 + \frac{1}{2}\sum\limits_{i=1}\limits^{p-1}K_{\rm{B}}(u_{i},v_{p}-u_{i})n(u_{i}),n(v_{p}-u_{i})  \nonumber\\
- n_p\sum\limits_{i=1}\limits^{N_{max}}K_{\rm{B}}(u_{i},v_{k})n(u_{i}) + \left. \frac{\partial n(v_{p})}{\partial t}\right |_{PC},
\end{eqnarray}
where $n(v_p)$ and $\Phi(v_p)$ are the number density and flux of volume $v_p$ particles, respectively,  $r$ is the planetocentric radius and $K_{\rm{B}}$ is the coagulation kernel that is dominantly controlled by the random collisions among particles (Brownian coagulation dominates over other processes for sub-micron size particles, which as we will see is the case for the resulting aerosol particle sizes). The first sum on the right hand side describes  the production of size $v_{p}$ particles from smaller particles, the second sum is the loss of size $v_{p}$ particles by coagulation to form bigger particles,  $\left.\partial n(v_{p})/\partial t \right |_{PC}$ describes the local photochemical production of this size particles (only for the first bin particles) and N$_{max}$ is the number of volume bins considered in the calculations. We use 30 bins with an expanding bin structure covering the range between 1 nm to $\sim$1 $\mu$m particle bulk radii (the radius of an equivalent mass spherical particle if the assumed shape is not spherical). The particle vertical transport is controlled by their gravitational settling velocity, and the influence of atmospheric mixing that is described through an eddy diffusion profile. The combined effect on the flux, $\Phi(v_p)$, of the particles is:
\begin{equation}
\Phi(v_p) = -V_s n(v_p) - K_{\rm ZZ}n\frac{\partial (n(v_p)/n)}{\partial r}
\end{equation}
where $V_s$ is the settling velocity of particles $p$ (see \cite{Lavvas10}), $K_{\rm{ZZ}}$ the eddy mixing profile and $n$ the atmospheric density. Below we discuss in detail specific parameters that affect the results of our calculations. 

\begin{figure}[!t]
\centering
\includegraphics[scale=0.5]{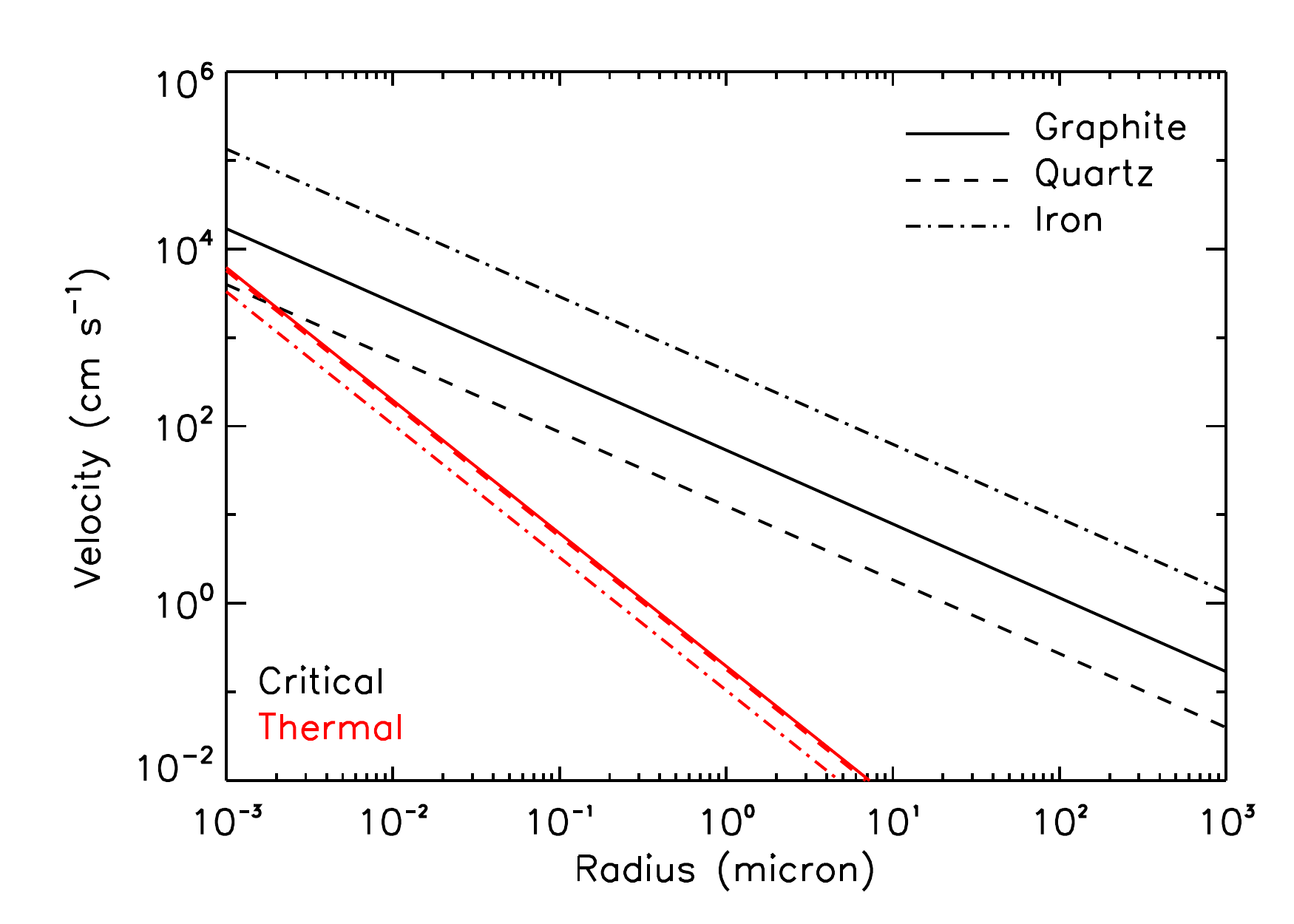}
\caption{Variation of critical velocity (black) and thermal particle velocity (red) at T=1000 K with particle radius, for different compositions. Parameters for the evaluation of the critical velocity for each composition are taken from \citep{Chokshi93}.}\label{stick}
\end{figure}

\subsection{Particle collisions}

Before evaluating the possible aerosol particle growth in exoplanet atmospheres we need to confirm that collisions among particles will permit their coagulation. Although this assumption is valid for the solar system studies where particle thermal velocities are small and collision energies are low, the high temperatures of EGPs and their resulting high thermal velocities, could limit the sticking efficiency among particles. The critical factor is the bond energy between the colliding particles that depends on the material they are made of. As the latter parameter is not known we can evaluate the bond strength under different composition assumptions and compare with the collision energies anticipated in exoplanet atmospheres. 

For this task we use the approach developed for collisions among dust particles in the interstellar medium \citep{Chokshi93,Dominik97} that is based on the surface energy and contact theory of solids \citep{Johnson71}. The approach utilises macroscopic parameters for the elastic properties of the particle's material such as Young's modulus, E, surface energy, $\gamma$, and Poisson ratio, $\nu$, and allows for the evaluation of the maximum collision velocity of two spherical particles for which the particles will stick together. Application of this approach to exoplanet conditions demonstrates that for an atmospheric temperature of  T=1000 K, particle collisions would allow sticking and thus growth for compositions similar to quartz, iron, and graphite (Fig.~\ref{stick}). These are characteristic examples of possible chemical compositions that could exist in EGP aerosols and provide an overall picture for the limits of the particle collision energy allowing their microphysical growth. Only for very small particle radii, below $\sim$3 nm, the critical and thermal velocities for quartz composition approach each other, which would imply that particles would have to reach this limiting radius through interaction with the gas phase (condensation or chemistry), before coagulation can permit further growth. However, at such small sizes the use of the macroscopic theory is questionable, while other theoretical studies demonstrate that at this size range long range interactions, such as van der Waals forces, are important and can significantly enhance the coagulation rates \citep{Harris88}. Temperature changes incur small changes to the above picture, thus we conclude that particle collisions in exoplanet atmospheres can result in their coagulation. 

\subsection{Particle ablation}
The model described above could be directly applied to the giant planets of our solar system, since it encompasses the main processes controlling the aerosol particle evolution in the pressure regions probed by observations. However, for EGPs we need to further consider the impact of the high temperatures observed in their atmospheres on the properties of the produced aerosol particles. For the gaseous species of the atmosphere, we know that high temperatures drive their thermal decomposition. A similar procedure can affect the aerosol particles when they reach to atmospheric regions with significantly warmer temperatures than those at their formation region. We can treat this process as an ablation procedure in which case the particles loose mass at a rate $\dot{m_p}$ that depends on the atmospheric temperature, the size of each particle, and the vapor pressure of their material. We included this mechanism in our calculations by adding to the continuity equation described previously one more term:
%\begin{equation}
%\frac{\partial n_p}{\partial t} = -\nabla\Phi_p + \left . \frac{\partial n_p}{\partial t}\right |_{Coagulation} + \left . \frac{\partial n_r}{\partial t}\right |_{Evaporation}
%\end{equation}
%with:
\begin{equation}
\left . \frac{\partial n_p}{\partial t}\right |_{Ablation} = -\frac{\partial{(\dot{m_p}n_p})}{\partial m} .
\end{equation}
This procedure is similar to the description of evaporating cloud particles and more details for the evaluation of all terms in the above equations can be found in \cite{Lavvas11c}.

\begin{figure}[!t]
\centering
\includegraphics[scale=0.5]{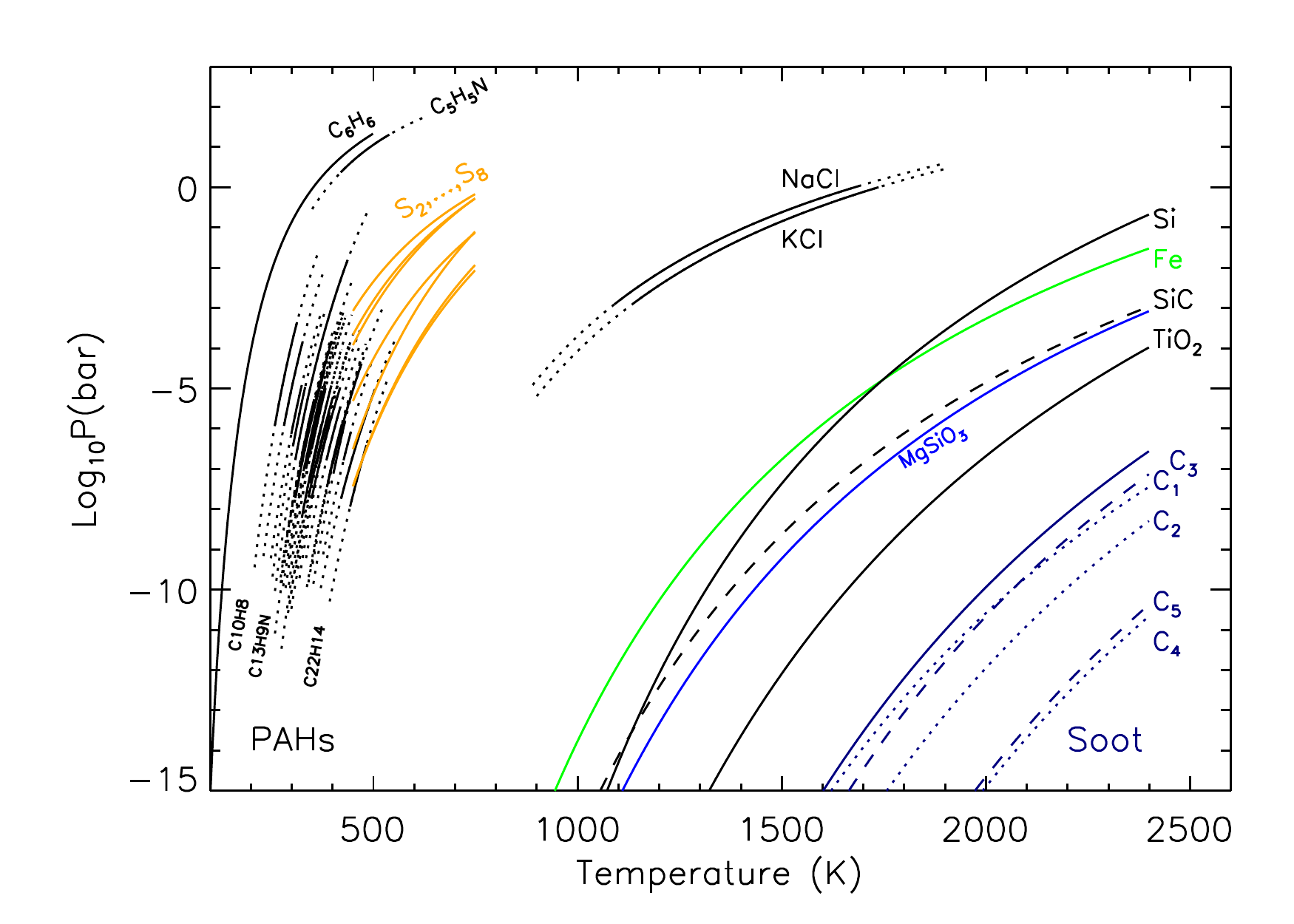}
\caption{Saturation vapor pressure curves of different compounds. Parameters for aromatic compounds are taken from \cite{Oja98} and \cite{Goldfarb08}, and for sulphur allotropes from \cite{Lyons08}. The Si (black) and Fe (green) curves present the saturation vapor pressure over liquid Si (NIST) and solid Fe \citep{Visscher10}, respectively. For enstantite (blue curve) we use the vapor pressure estimated in \cite{Cameron85}. Parameters for SiC and TiO$_2$ are taken from \cite{Lilov93} and \cite{Woitke04}, respectively. For soot (navy curves), saturation vapor pressures are from \cite{Michelsen07} with the different broken lines presenting the contribution of various carbon clusters. NaCl and KCl curves are from NIST.}\label{vpdata}
\end{figure}

The importance of aerosol ablation depends on the chemical composition of the particles, which controls their resistance at high temperatures. Figure \ref{vpdata} demonstrates this sensitivity by presenting the saturation vapor pressure curves of different compounds. Included are characteristic examples of polycyclic aromatic compounds, which are believed to contribute to the photochemical aerosols of Jupiter and Saturn \citep{Wong00,Wong03} and the aerosols in Titan's atmosphere \citep{Lavvas11a}, while they are also observed in the soots formed from combustion engines \citep{Frenklach02}, as well as in the interstellar medium dust \citep{Draine03}. These are mainly based on carbon (and nitrogen for Titan and ISM) chemistry. However, the hot environments of exoplanets are potentially rich in more exotic species such as S, Si, Ti, and other metals that could also lead to the production of complex molecules and eventually form aerosol particles. For example, silicon species such as silane (SiH$_4$) and its photochemical products have a similar chemical complexity with the more familiar (from the atmospheric chemistry point of view) hydrocarbons derived from the photolysis of methane, while sulphur compounds have already been proposed as possible components of exoplanetary aerosols \citep{Zahnle16}. 

Silicon carbide (SiC) is another candidate that is usually found in carbon rich stars, while other silicate mixtures such as enstantite (MgSiO$_3$), forsterite (Mg$_2$SiO$_4$), quartz (SiO$_2$), and oxidies of titanium or vanadium (TiO$_2$, VO) are usually considered as possible contributors to the formation of condensates in the deeper regions of exoplanet atmospheres \citep[e.g.][]{Lecavelier08}, and we include them here for comparison. Recent studies on the physical properties of such condensates demonstrate that they can not reproduce well the observed primary transit spectra for HD 189733 b \citep{Lee15, Pinhas17}, although they can satisfy secondary eclipse constraints \citep{Lee17}. These results indicate that a different chemical composition heterogeneous compound is required in the upper atmosphere, relative to the dominant silicate condensates in the lower atmosphere. 

Among the compounds we investigated, the material with the highest resistance to extreme temperatures is soot. Although this term encompasses a large family of combustion or pyrolysis products of carbon chemistry, all characteristic examples we found have the lowest rate of evaporation at high temperatures. Soot is a common product of high temperature chemistry, but its formation in a planetary atmosphere will depend on the availability of carbon species. A soot composition haze has been suggested for the cooler atmosphere (T$_{eff}$$\sim$550 K) of GJ 1214 b \citep{Morley13}. Here we assume a soot composition and explore the possibility of such a photochemical aerosol formation in hot exoplanets and the resulting particle properties. 

\subsection{Aerosol production}
% Aerosol production rates - photochemical model results
In order to initiate the simulation of aerosol evolution we need to assume an aerosol production profile that generates particles in the first bin of the aerosol size grid. If we assume that the production profile can be represented by a gaussian distribution in pressure then we need to specify the location of the peak, the width of the distribution, and the magnitude of aerosol production, i.e. the mass production rate of aerosols. Although we do not have any observational constraints for these properties we can derive some indicative values based on photochemical models and what we know from the solar system. Taking as example the atmosphere of Jupiter, the mass flux of aerosols inferred from observations ($\sim$7$\times$10$^{-14}$ g cm$^{-2}$s$^{-1}$) corresponds to about 1$\%$ of the total mass flux generated by the photolysis of methane \citep{Moreno96,Wong00}. The ion-neutral processes leading to the formation of photochemical aerosols in Titan's upper atmosphere have a similar efficiency, although the total mass flux of aerosols is further increased through neutral chemical processes and reaches about 10$\%$ of the mass flux generated by the photolysis of Titan's major atmospheric compounds, N$_2$ and CH$_4$ \citep{Lavvas13}. Thus, we can start our simulations assuming a similar range of efficiencies for the production of aerosols in exoplanets. 
%production profile
The location of the aerosol production profile can also be estimated from the photochemical model results. These suggest that most of the high energy photons are deposited in the upper atmosphere, above the 1 $\mu$bar level. Hence, we center our aerosol production profile at this location, and assume a profile width that ranges between 0.1 and 10 $\mu$bar. 

\subsection{Particle shape}
The shape of the particles depends on the phase of their material (liquid/solid), as well as their interaction with the gas phase background; coagulation of liquid particles results in the formation of a larger spherical particle, while for solid particles collisions lead to their aggregation. However, in the latter case, heterogeneous processes at the surface of the particles (e.g. chemical reactions, adsorption, condensation) can modify aggregates towards a quasi-spherical shape, if the mass added heterogeneously from the gas phase is comparable to the mass added by coagulation \citep{Mitchell03, Morgan07, Lavvas11c,Lavvas11a}. Thus, the theoretical evaluation of the particle shape requires an in depth knowledge for the mechanism(s) generating the particles, which is lacking at the current level of aerosol investigation for exoplanet atmospheres. We perform our calculations assuming spherical particles, and we subsequently conclude that aggregate structures are not very likely in the upper atmosphere but may form in the lower atmosphere where conditions for aggregation are more favorable.

\subsection{Particle charge}
Another parameter that affects the growth of particles is the potential charge on their surface. Collisions among charged particles can increase or decrease the resulting particle size, if the particles have a different or same sign charge, respectively. Typically, the effect of particle charging is taken into account through a charge density parameter, $\chi$, which describes the total number of charges, $Z$, on the surface of a particle as a function of its bulk radius, $r$, ($Z=\chi r$). Subsequently, the total charge of the particles is used for the calculation of the particle sticking efficiency, $\alpha$ \citep[e.g.][]{Lavvas10}.
%through:
%\begin{equation}
%\alpha \simeq \exp(-\frac{Z_1Z_2}{k_BT(r_1 + r_2)})=\exp(-\frac{r_1r_2\chi^2}{k_BT(r_1 + r_2)})
%\end{equation}
%where we have assumed the general case of two different particles with radii $r_1$ and $r_2$  \citep[e.g.][]{Lavvas10}. 
Typical values for $\chi$ range between 10 and 30 e$^{-}$$\mu$m$^{-1}$ in the solar system aerosols, which means that electrostatic effects start to modify the particle growth for sizes above 1 $\mu$m, for the typical temperatures in the solar system. For the higher temperatures found in EGPs the corresponding effect of charging would be significantly reduced if $\chi$ had a similar magnitude. However, charging rates for extrasolar planets could be significantly different from those in our solar system due to the presence of high densities of free electrons from the photoionization of Na and K \citep{Lavvas14}, and due to the possibly higher impact of the photoelectric effect on particle ionization, resulting from the enhanced stellar fluxes in EGP environments. In our calculations we assume that the charging effects are negligible ($\chi$=0), and subsequently demonstrate that this assumption is valid for the upper atmosphere probed by the transit observations, due to the small size of the simulated particles that is controlled by other processes.

\section{Case study}

\begin{figure}[!t]
\centering
\includegraphics[scale=0.5]{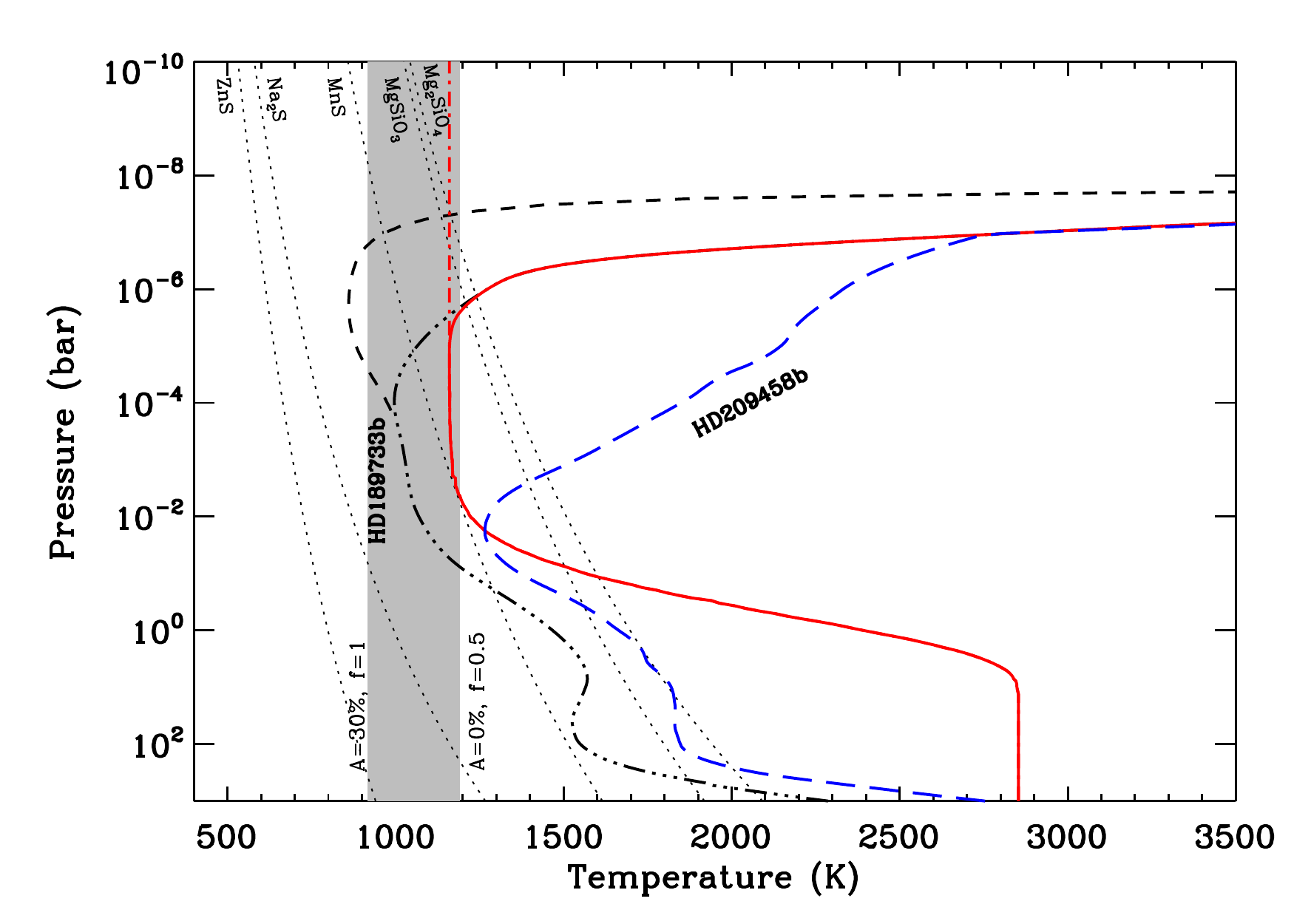}
\caption{Temperature profiles for the atmospheres of HD 209458 b (blue, long-dashed line, MK case) and HD 189733 b (black dashed line, M profile; black dash-triple-dotted line, MK profile; red solid line, LK profile; red dashed-dotted line, L profile, see text for details) corresponding to day average illumination conditions. The grey band presents the range of skin temperatures anticipated for HD 189733 b, bound by a high bond albedo case (A=30$\%$) with complete redistribution of incoming radiation (f=1), and a low albedo case (A=0$\%$) with no redistribution to the night side (f=1/2). Dotted lines present the conditions for the formation of condensates ZnS, Na$_2$S, MnS, MgSiO$_3$, and  Mg$_2$SiO$_4$, assuming thermochemical equilibrium everywhere in the atmosphere and solar metallicity \citep{Visscher10}.}\label{pT}
\end{figure}

We proceed now to the application of the model to HD 189733 b and HD 209458 b. We use the \cite{Lavvas14} exoplanet chemistry kinetics model for the simulation of the atmospheric composition from which we derive the potential aerosol mass fluxes for each planet {\color{\clr}(see \cite{Vardavas07} and \cite{Lavvas08} for more details on the chemistry model)}. The background chemistry depends strongly on the assumed elemental abundance of each system, on the stellar insolation reaching the planetary atmosphere, on the atmospheric thermal structure, and on the efficiency of the atmospheric mixing. We consider solar elemental abundances in our calculations. For the stellar flux we use a compilation based on the $\epsilon$-Eridani spectrum that has a similar spectral type to HD 189733 b and a Phoenix model of appropriate parameters for this star. For HD 209458, we use the solar spectrum \citep[see][]{Lavvas14}. Current temperature profiles suggested for HD 189733 b include the studies by \cite{Line10,Line14} and \cite{Moses11}. Here we only consider the two limiting cases of the later two studies. The \cite{Moses11} p-T profile (corresponding to day averaged conditions and designated as M profile below) is based on a GCM simulation from \cite{Showman09} smoothly connected to un upper atmosphere escape model \citep{Yelle04}. The \cite{Line14} profile is retrieved from secondary eclipse observations. These authors report a p-T profile in the range between 1 mbar to 10 bar, with the retrieval primarily sensitive to the pressure range 10-100 mbar. In order to extrapolate to other pressures in our calculations we assumed either an isothermal behaviour in both higher and lower pressure regimes (designated as L profile), or an isotherm profile to higher pressures and a smooth connection to the escape calculations of \cite{Koskinen13a} for the upper atmosphere (designated as LK). We also consider another temperature profile (MK) that utilises the cooler temperature p-T profile from the GCM calculations smoothly connected to the upper atmosphere thermal structure calculated by \cite{Koskinen13a}. The MK temperature profile is in better agreement with the skin temperatures anticipated for this atmosphere (Fig.~\ref{pT}). Finally for the comparison with HD 209458 b we assume the corresponding day average p-T profile from \cite{Moses11} for this atmosphere. For the atmospheric mixing we use the K$_{\rm ZZ}$ profiles suggested by GCM models for the two planets, but we perform sensitivity test for 0.1$\times$K$_{\rm ZZ}$ and 0.01$\times$K$_{\rm ZZ}$ profiles. Below we discuss the results for the chemical composition and estimates of the aerosol mass fluxes, the resulting properties of the formed aerosols, and their impact on the planetary transit.

\begin{figure*}[!t]
\centering
\includegraphics[scale=0.5]{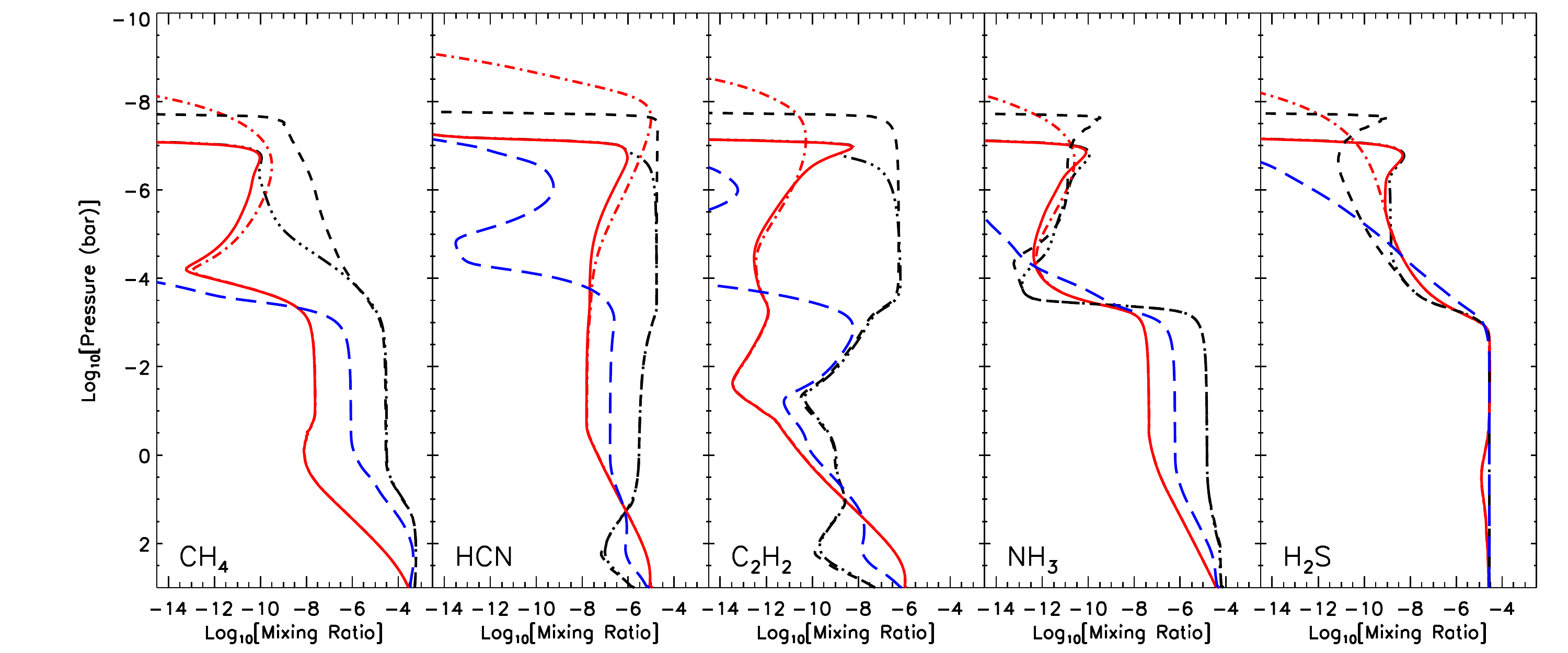}
\includegraphics[scale=0.5]{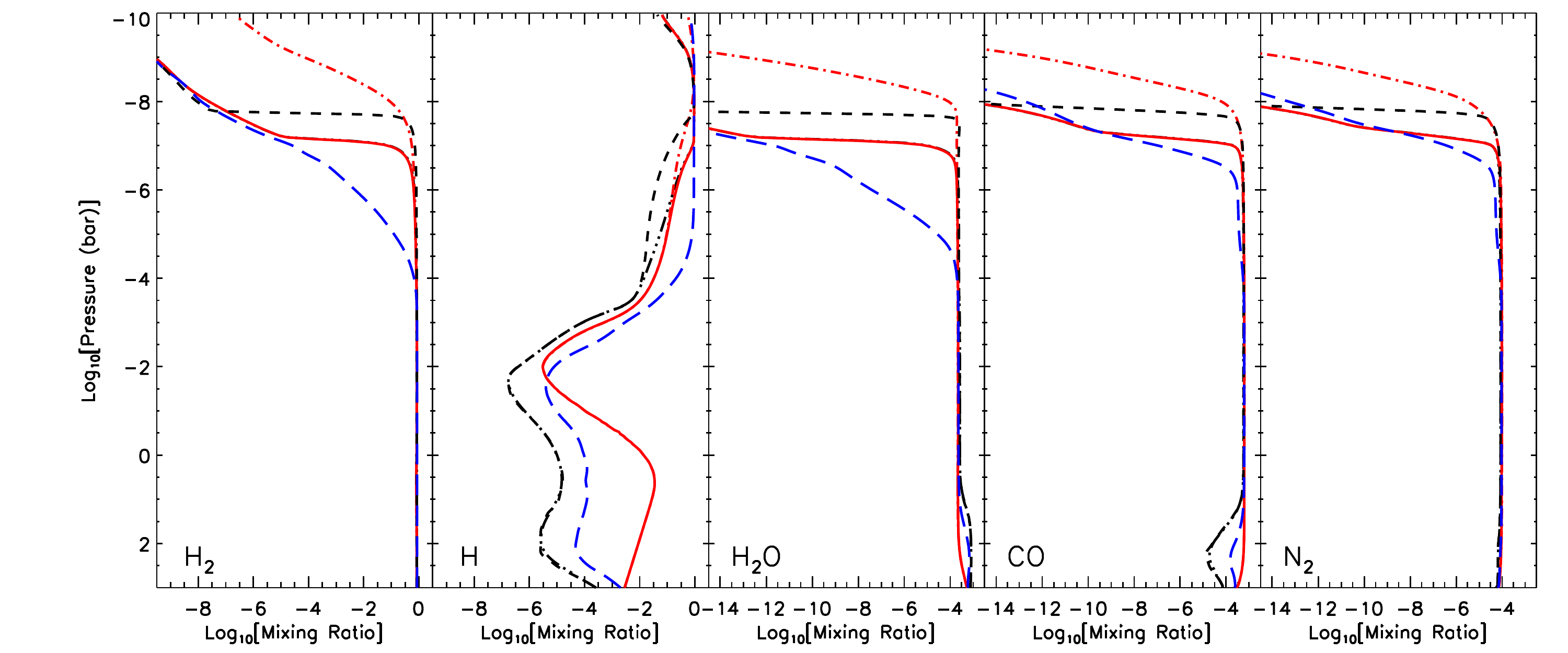}
\caption{Mixing ratio profiles of major species for the different temperature profiles assumed for HD 2093458 b and HD 189733 b. We assumed the nominal K$_{\rm ZZ}$ profile for this simulation. Lines styles and colors correspond to the temperature profiles of Fig.~\ref{pT}, i.e. red solid: LK profile, red dash-dotted L profile, black dashed M profile, black dsh-triple-dotted MK profile, blue long dashed HD 209458 b's profile.}\label{chem}
\end{figure*}

\subsection{Photochemical mass fluxes}

In order to evaluate the significance of aerosol production, we first compare the mass fluxes generated by the photolysis of major compounds in the atmospheres of HD 209458 b and HD 189733 b, under the different thermal structures assumed (Table~\ref{mflux}). The mass flux from the photolysis of methane, or of its primary products (HCN, C$_2$H$_2$), is significantly larger in the atmosphere of HD 189733 b relative to HD 209458 b. In both atmospheres and under all assumed temperature profiles, methane is destroyed near $\sim$1 mbar and is replaced primarily by HCN and to a second degree by C$_2$H$_2$. However, the abundance of these species surviving in the upper atmosphere is significantly larger for HD 189733 b compared to HD 209458 b (Fig.~\ref{chem}). This difference is the result of the rapid temperature increase at p$<$10 mbar in HD 209458 b (Fig.~\ref{pT}) that drives the atmospheric chemical equilibrium closer to the thermochemical limit, i.e. the carbon balance is efficiently shifted towards CO and the CH$_4$ abundance surviving in the upper atmosphere is significantly reduced on HD 209458 b, compared to HD 189733 b. In practical terms, the combined mass flux from the photolysis of CH$_4$ and of its secondary products is between 2 and 4 orders of magnitude smaller in HD 209458 b relative to that of HD 189733b, depending on the assumed temperature profile for the latter atmosphere. Thus, we note that the thermal structure in the transition between the lower and upper atmosphere (practically the pressure range between 1 mbar to 1 $\mu$bar) has a major impact on the atmospheric composition from the point of view of complex chemical products, and current observations do not provide solid constraints for this parameter on either planet.

The large differences observed between the two planets considered should not obscure the sensitivity of the composition results on the assumed temperature profiles for the HD 189733 b atmosphere. Evidently the large temperature difference between the simulated profile (M) and the retrieved profile (L) in the lower atmosphere (more than 1000 K, see Fig.~\ref{pT}) results in changes for the chemical composition of the whole atmosphere (Fig.~\ref{chem}), which could affect the primary and secondary eclipse observations if the accuracy of the measurements is improved in the near future. This variability is also evident in the calculated mass fluxes (Table~\ref{mflux}). The total mass flux from the photolysis of CH$_4$, C$_2$H$_2$, and HCN ranges between 2.2$\times$10$^{-12}$ to 9.7$\times$10$^{-11}$ g cm$^{-2}$s$^{-1}$ among the four temperature profiles considered. The relative variation for the individual molecules can be much larger. Assuming an 1$\%$ efficiency for aerosol formation results in aerosol mass fluxes of the order of 10$^{-14}$-10$^{-12}$  g cm$^{-2}$s$^{-1}$, which are large enough to have important ramifications for the resulting aerosol opacity. The question arising from this comparison is why the simulated and retrieved conditions are so different and how the simulated p-T profile could be brought in better agreement with the observations? Could the heating by the anticipated hazes help to this aspect? {We evaluate this scenario further below.}

\begin{table*}[!t]
\centering
\caption{Mass fluxes (in g cm$^{-2}$s$^{-1}$) produced from the photolysis of different species in the upper atmosphere (p$\le$10$^{-5}$ bar) of HD 209458 b and of HD 189733 b under different assumptions for the thermal structure. a(b) = a$\times$10$^b$. }\label{mflux}
\begin{tabular}{r|ccc|ccc}
\hline
\hline
			& \multicolumn{6}{c}{HD 189733 b} \\
			& \multicolumn{3}{c}{M}					&  \multicolumn{3}{c}{MK} \\	
Species		 & K$_{\rm ZZ}$ & 0.1$\times$K$_{\rm ZZ}$ & 0.01$\times$K$_{\rm ZZ}$ & K$_{\rm ZZ}$ & 0.1$\times$K$_{\rm ZZ}$ & 0.01$\times$K$_{\rm ZZ}$\\	
\hline
CH$_4$		&8.1(-14)	&7.1(-17) & 1.9(-18)		&7.1(-16) &2.2(-19)\rm	& 3.7(-21)\\
HCN			&6.4(-11)	&2.8(-11) & 2.6(-12)		&2.9(-11) &2.0(-12)\rm	& 6.8(-14)\\
C$_2$H$_2$	&3.3(-11)	&6.5(-15) & 1.9(-19)		&1.8(-11) &8.1(-16)\rm	& 4.0(-18)\\
CO			&1.5(-12)	&2.3(-13) & 5.0(-16)		&4.8(-12) &2.0(-12)\rm	& 4.9(-15)\\
N$_2$		&8.3(-13)	&4.2(-13) & 6.0(-12)		&7.0(-13) &3.3(-13)\rm	& 4.2(-14)\\
NH$_3$		&1.2(-15)	&1.9(-16) & 2.7(-15)		&1.6(-15) &4.4(-17)\rm	& 8.3(-18)\\
H$_2$S		&3.0(-14)	&9.1(-15) & 2.1(-14)		&5.9(-13) &9.3(-14)\rm	& 5.6(-14)\\
S$_3$		&6.8(-12)	&2.5(-12) & 8.5(-11)		&1.4(-12) &4.0(-13)\rm	& 3.0(-13)\\
\it Soot (1$\%$)	&\it 9.7(-13)	& \it 2.8(-13)	& \it2.6(-14)		&\it 4.7(-13)	& \it 2.0(-14) 		& \it 6.8(-16)\\
\hline
			& \multicolumn{3}{c}{L}	& \multicolumn{3}{c}{LK}\\
Species		& K$_{\rm ZZ}$ & 0.1$\times$K$_{\rm ZZ}$ & 0.01$\times$K$_{\rm ZZ}$ & K$_{\rm ZZ}$ & 0.1$\times$K$_{\rm ZZ}$ & 0.01$\times$K$_{\rm ZZ}$\\	
\hline
CH$_4$		 &1.0(-15)	&4.6(-20) 	& 1.6(-22)		&2.7(-16) &1.3(-19) & 2.1(-21) \\
HCN			 &2.9(-11)	&1.4(-11) 	& 7.7(-15)		&2.2(-12) &2.9(-13) & 1.2(-14) \\
C$_2$H$_2$	 &1.5(-15)	&9.7(-19)	& 1.9(-22)		&3.0(-14) &5.5(-16) & 4.3(-18) \\
CO			 &8.5(-13)	&3.3(-13) 	& 4.7(-17)		&5.1(-12) &2.3(-12) & 6.1(-15) \\
N$_2$		 &9.7(-13)	&4.1(-13) 	& 6.3(-12)		&7.7(-13) &3.7(-13) & 6.1(-12) \\
NH$_3$		 &8.2(-16)	&7.0(-18) 	& 1.6(-16)		&6.4(-16) &6.0(-18) & 1.5(-16) \\
H$_2$S		 &4.0(-13)	&1.1(-13) 	& 9.0(-14)		&5.0(-13) &1.2(-13) & 9.2(-14) \\
S$_3$		 &1.6(-13)	&5.0(-14) 	& 4.3(-14)		&1.5(-13) &5.0(-14) & 4.3(-14) \\
\it Soot (1$\%$)	 &\it 2.9(-13)&\it 1.4(-13) & \it 7.7(-17)		&\it 2.2(-14) & \it 2.9(-15) & \it 1.2(-16) \\
\hline
\end{tabular}
\begin{tabular}{r|ccc}
\hline
			& \multicolumn{3}{c}{HD 209458 b} \\
Species		& K$_{\rm ZZ}$ & 0.1$\times$K$_{\rm ZZ}$ & 0.01$\times$K$_{\rm ZZ}$ \\	
\hline

CH$_4$		& 1.1(-22) & 7.5(-25)\rm	& 4.7(-25)\\
HCN			& 7.8(-15)	& 2.3(-15)\rm	& 3.2(-15)\\
C$_2$H$_2$	& 2.8(-17)	&1.2(-18)\rm	& 2.3(-18)\\
CO			& 4.8(-11)	&3.0(-11)\rm	& 1.7(-11)\\
N$_2$		& 4.5(-12)	&2.2(-12)\rm	& 9.9(-13)\\
NH$_3$		& 3.5(-16)	&9.3(-17)\rm	& 9.2(-17)\\
H$_2$S		& 1.8(-11)	&7.2(-12)\rm	& 6.8(-12)\\
S$_3$		& 1.8(-19)	&9.2(-20)\rm	& 9.0(-20)\\
\it Soot (1$\%$)	 &\it 7.8(-17)&\it 2.3(-17) & \it 3.2(-17) \\
\hline
\hline
%\multicolumn{3}{l}{a(b) = a$\times$10$^b$}\\
\end{tabular}
\end{table*}

\begin{figure}
\centering
\includegraphics[scale=0.5]{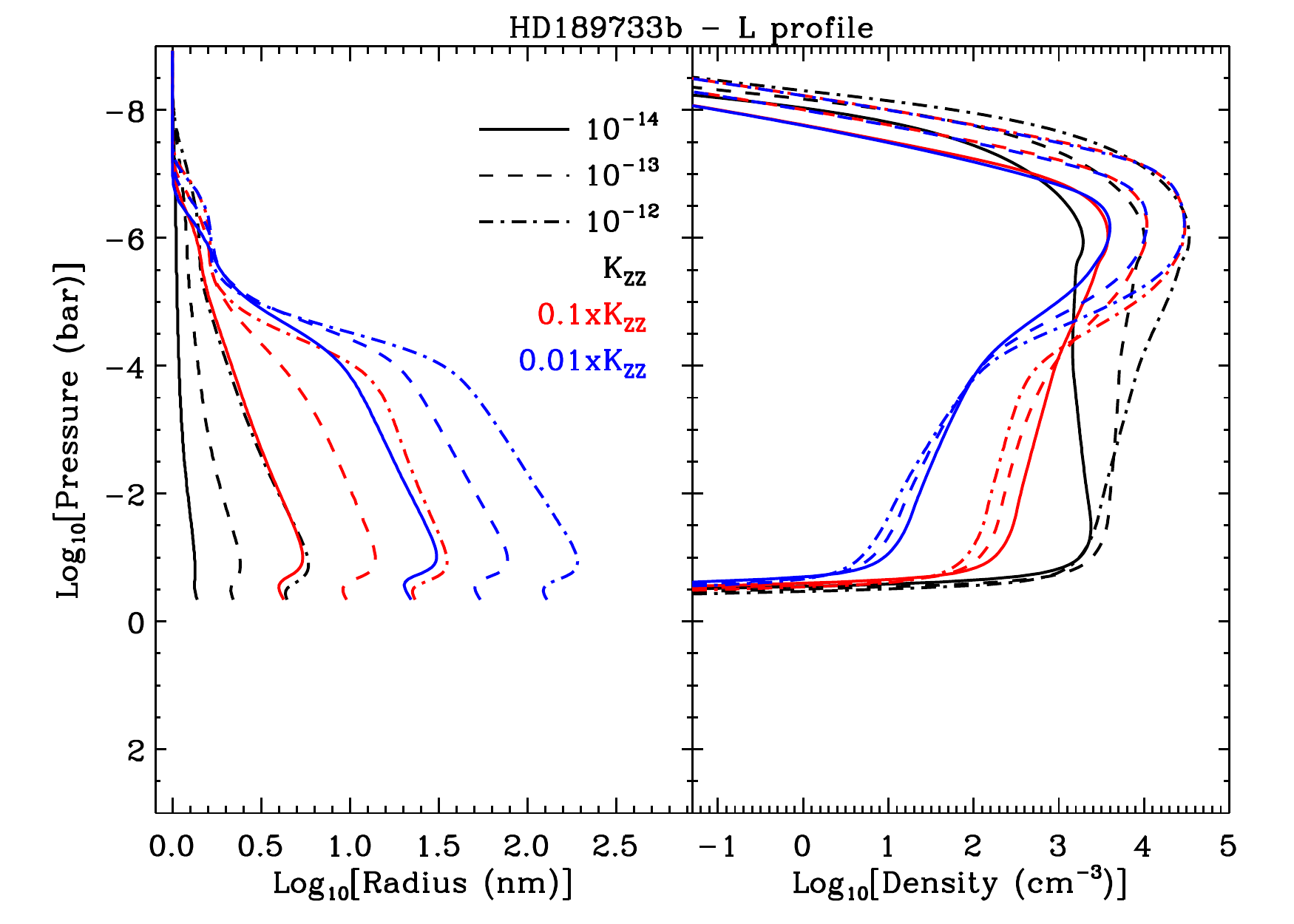}
\includegraphics[scale=0.5]{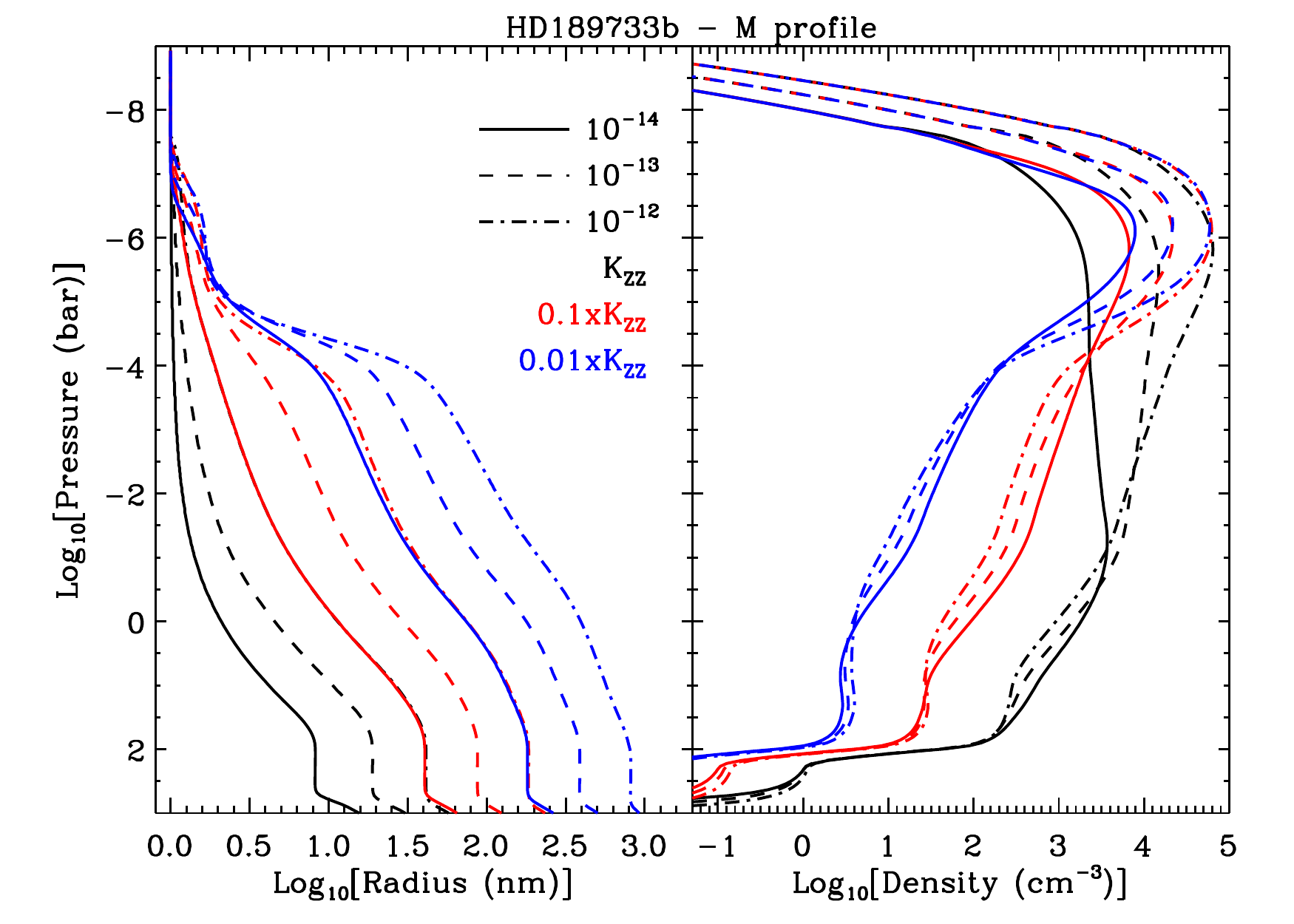}
\includegraphics[scale=0.5]{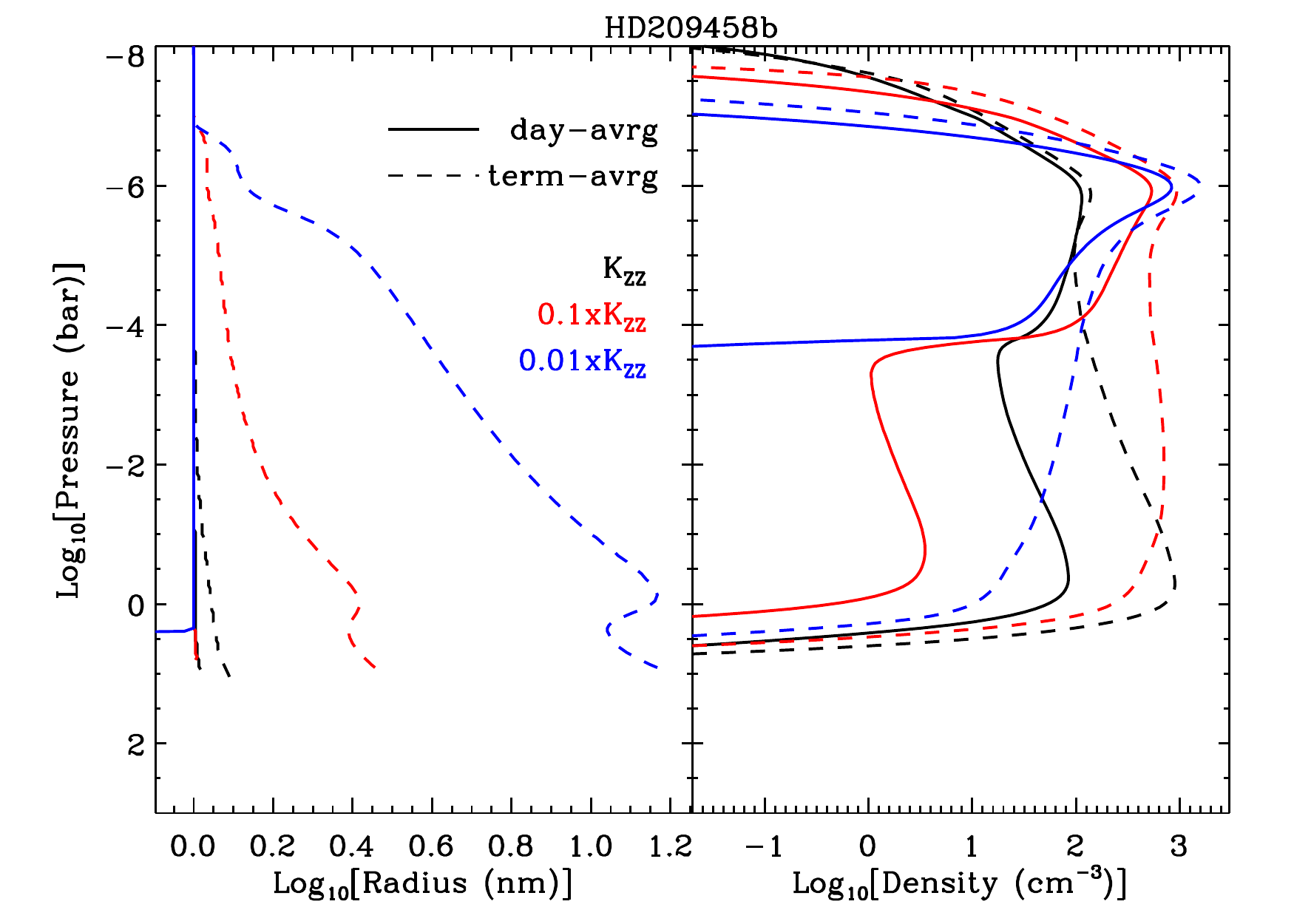}
\caption{Average aerosol particle radius and corresponding density, under different assumptions for the aerosol mass flux (line style) and atmospheric mixing (colors). The first two panels present results for the case of HD 189733 b for the L and M temperature profiles, while the last panel presents the model results for the case of HD 209458 b.}\label{micro}
\end{figure}

%Before continuing with the evaluation of the physical properties of the aerosols potentially formed in the atmosphere of HD 189733 b, 
We further note that the differences in the thermal structure assumed in the upper atmosphere of HD 189733 b have further ramifications for the physical processes taking place there. Comparison of the model results between the (M, MK) or the (L, LK) profiles demonstrate how the composition is modified assuming changes only in the thermal structure of the upper atmosphere (see Fig.~\ref{pT}). Lower temperatures allow for the survival of species to higher altitudes (lower pressures). This characteristic is particularly evident when comparing the L profile with the LK profile where the latter includes a hot thermosphere suggested by atmospheric escape models. The escape models indicate that the temperature should increase within a few atmospheric scale heights as soon as the dominance of H$_2$ is overthrown by the increasing abundance of atomic hydrogen \citep{Yelle04,GarciaMunoz07,Koskinen07, Koskinen13a}, which is formed through the catalytic destruction of H$_2$ by OH \citep{Moses11}, as well as, by S and O (which in turn are produced from the destruction of their parent molecules, H$_2$O, H$_2$S). The increasing H abundance has its own ramifications to the abundances of different species, which depend on the chemical pathways defining the chemical equilibrium for each case, but in general leads to the destruction of molecular structures (Fig.~\ref{chem}). At lower temperatures the H$_2$ to H transition is slower, thus molecular species can survive at higher altitudes. As the production of photochemical aerosols depends critically on the production of large molecular structures, the temperature assumed in the transition between the lower atmosphere and the thermosphere can play an important role in the nature and abundance of photochemical aerosols.

Apart from the temperature, assumptions on the eddy mixing profile have an impact on the simulated composition profiles, therefore on the resulting aerosol mass fluxes. We consider as nominal mixing profile for each planet case the profiles presented in \cite{Moses11}, which are based on GCM simulations \citep{Showman09}. Subsequent studies suggested that the actual atmospheric mixing efficiencies could be lower depending on the way the eddy diffusivity is calculated based on the GCM simulations \citep{Parmentier13}. In order to evaluate the possible influence of a reduced atmospheric mixing we also performed calculations with 10$\times$ and 100$\times$ reduced eddy profiles relative to the nominal case (see Table~\ref{mflux}). This evaluation demonstrates that with reduced atmospheric mixing the simulated mass fluxes are reduced as anticipated; for CH$_4$ the reduced eddy profile allows its abundance to remain close to the thermochemical equilibrium solution at lower pressures than for the nominal eddy profile case, thus resulting in a smaller methane abundance in the upper atmosphere. On the other hand, the mixing ratios of CO, N$_2$ and H$_2$O  are controlled by thermochemical equilibrium at high pressures and remain constant for the rest of the atmosphere before diffusive separation decreases their abundance in the upper atmosphere \citep{Moses11}. Thus, reduced mixing for these species lowers their homopause altitude, reducing in this way their abundance in the upper atmosphere. However, for all cases considered HCN remains the most abundant hydrocarbon in the upper atmosphere and its photolysis provides significant mass fluxes for the formation of hazes.

Finally, other species also generate significant mass fluxes of higher order compounds. For example the chemistry of H$_2$S leads to the formation of  poly-sulfur compounds, from which we track species up to S$_3$ in these simulations. The mass flux from the photolysis of S$_3$ is also significant ranging between $\sim$10$^{-13}$ to $\sim$7$\times$10$^{-12}$ g cm$^{-2}$s$^{-1}$ for the different temperature profiles assumed on HD 189733 b, while it is more than 5 orders of magnitude smaller for the atmosphere of HD 209458 b. However, the high temperature conditions do not allow the sulfur compounds to condense in the simulated atmospheres. 

\subsection{Aerosol properties}

With the estimated mass fluxes we can evaluate the resulting properties of the aerosol particle distribution. We present simulations only for the M and L temperature profiles as the other cases only correspond to changes in the upper atmosphere that will not significantly modify the results of the aerosol microphysics occurring at lower altitudes, unless the temperature increases rapidly enough to affect the ablation of the aerosol particles. Figure~\ref{micro} presents an overview of our results in terms of the average particle size and corresponding density under different assumptions of the particle mass flux generated by the photochemistry, the eddy mixing efficiency, and the atmospheric thermal structure of HD 189733 b and HD 209458 b. For these simulations we assume that the initial particles formed by the photochemistry have a typical radius of 1 nm, therefore a given mass flux translates to a production rate of particles in the first bin of our aerosol grid. As soon as they are formed particles start to coagulate and grow in size, forcing the particle population to decrease. The efficiency of particle coagulation depends directly on the particle thermal velocity, i.e. the temperature, and the sticking efficiency of the particles (which for the moment we assume it is one for this evaluation, i.e. no charge effects). 

The differences in the two thermal profiles we consider for the upper atmosphere of HD 189733 b cause only small variations on the particle coagulation rates (which scale as the square root of the temperature), thus the resulting average particle size and density are similar for the corresponding mass fluxes and eddy profiles (Fig.~\ref{micro}). However, the temperature structure assumed in the lower atmosphere (below $\sim$10 mbar) has major implications for the resulting particle properties. For the L profile, assuming a soot type composition, particles thermally decompose below the 0.1 bar leading to their rapid loss, as demonstrated by the particle density profiles (Fig.~\ref{micro}). On the contrary for the cooler M profile, particle decomposition occurs for p$>$100 bar, and particles can survive to deeper pressure levels in the atmosphere where they grow to larger sizes. Assuming a different chemical composition for the photochemical aerosols, results in particle decomposition at cooler temperatures, i.e. at even lower pressures than for the soot composition case. 

%For example considering a composition based on sulphur results to particle destruction below XX bar for the M profile, while for the L profile particle should not form at all.

\begin{figure*}
\centering
\includegraphics[scale=0.5]{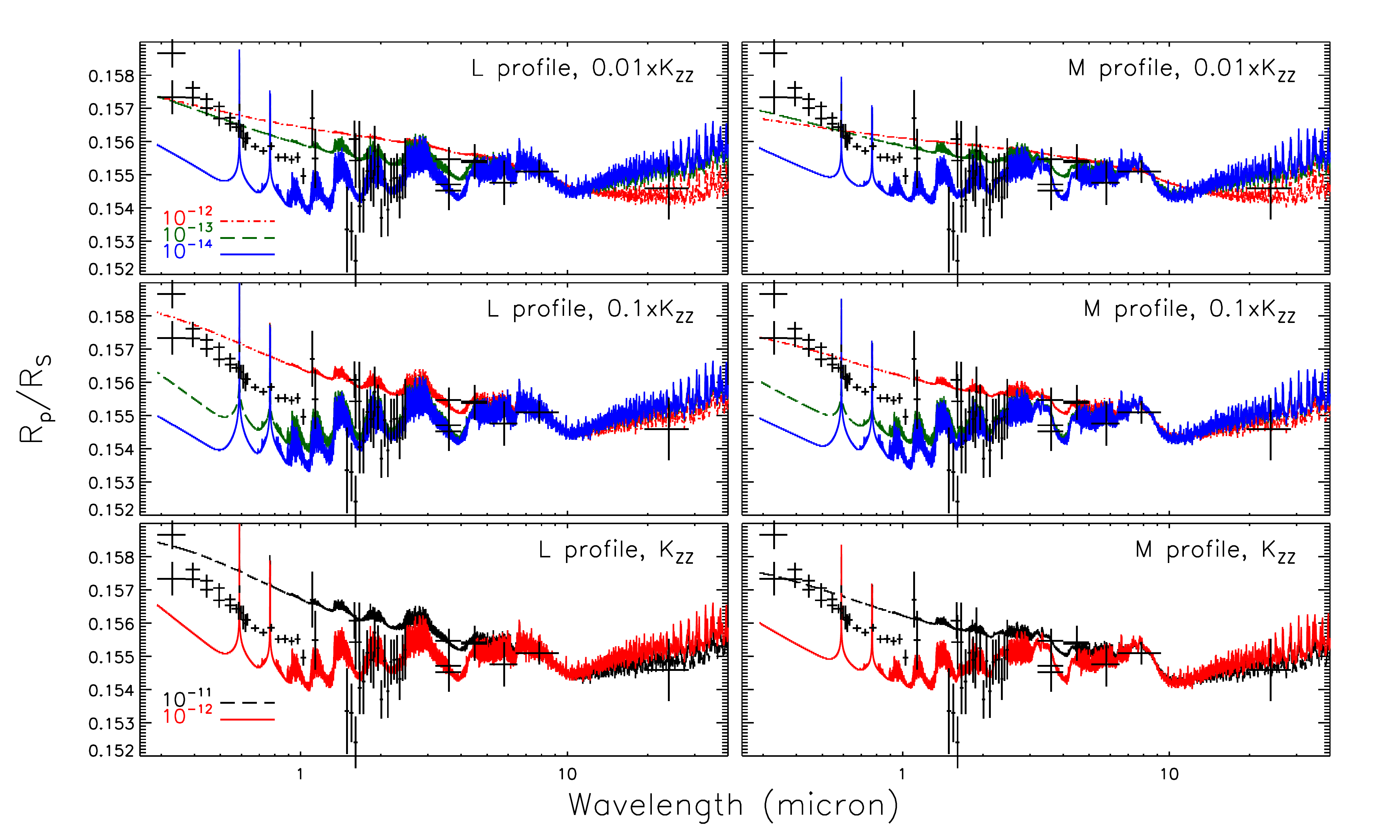}
\caption{Simulated $\rm R_P/R_S$ spectra of HD 189733 b (color lines) compared to observations (crosses) compiled by \cite{Pont12}. Each panel present results for a different combination of temperature profile and eddy mixing efficiency as designated. The colored lines present different aerosol mass flux cases with red, green, and blue corresponding to 10$^{-12}$, 10$^{-13}$, and 10$^{-14}$ g cm$^{-2}$s$^{-1}$, respectively.}\label{transit1}
\end{figure*}

The average particle radius depends on the mass flux assumed as this parameter defines the population of particles available for collisions, with larger values resulting {\color{\clr}in} larger particles: for the L profile, in the 10$^{-12}$ g cm$^{-2}$s$^{-1}$ mass flux case particles grow up to $\sim$6 nm, in the 10$^{-13}$ g cm$^{-2}$s$^{-1}$  case the average size reaches to 2.5 nm, while for 10$^{-14}$ g cm$^{-2}$s$^{-1}$ case particle coagulation is practically inefficient and the particle density profile demonstrates the effect of rapid transport from the formation region to the lower atmosphere where their population increases due to the slower vertical transport. For the M profile, aerosols grow further in the lower atmosphere, reaching maximum average radii of 8 nm, 20 nm, and 40 nm for the three mass flux cases considered.

Particle vertical transport has an indirect effect on their coagulation that dominates over the direct temperature influence. If particles are transported faster than they collide at a given pressure level their coagulation will be limited and thus their size will remain approximately constant until they reach a pressure region where collisions are efficient. Comparison of the model results at different eddy mixing efficiencies demonstrates this effect: for the nominal K$_{\rm ZZ}$ atmospheric transport limits the collisions of the formed particles and the resulting average particle size remains small and only increases in the deep atmosphere as the eddy mixing is reduced. Under the reduced atmospheric mixing conditions, collisions are more frequent and particles grow rapidly below their production region. For the L profile, particles reach average radii of $\sim$35 nm and $\sim$200 nm, for the 0.1$\times$K$_{\rm ZZ}$ and 0.01$\times$K$_{\rm ZZ}$ cases, respectively, under the highest mass flux case considered (10$^{-12}$ g cm$^{-2}$s$^{-1}$). The corresponding radii for the M case are, $\sim$185 nm and $\sim$750 nm. Rapid coagulation decreases the corresponding particle densities below their formation region, and results in progressively smaller densities in the lower atmosphere for the higher mass flux assumed (Fig.~\ref{micro}). A similar effect would result from the planet gravity field with larger mass planets pulling particles more efficiently, thereby reducing their growth, under otherwise similar temperature, particle mass flux, and atmospheric mixing conditions. 

The average particle properties for the atmosphere of HD 209458 b are strongly affected by ablation as the rapid temperature increase above the 10 mbar pressure level results in the partial decomposition of the particles formed at higher altitudes (Fig.~\ref{micro}). To further demonstrate the small contribution that photochemical hazes should have in the opacity of this atmosphere, we consider an aerosol mass flux of 10$^{-15}$ g cm$^{-2}$s$^{-1}$ that corresponds to $\sim$10$\%$ efficiency of photochemical aerosol formation for the largest photochemical mass fluxes generated according to Table~\ref{mflux}. For both the nominal and the reduced atmospheric mixing profiles the formed particles do not grow and only the population of the first radius bin is modified depending on how efficiently particles are moved away for parts of the atmosphere where they can decompose.  We also evaluated the resulting aerosol properties assuming the terminator average thermal structure reported by \cite{Moses11}. This scenario assumes that the formed photochemical aerosols are rapidly transported horizontally from their formation region to the limbs of the planet where the low temperatures allow for their survival. Under this assumption, particles sustain a larger population but their size does not significantly increase as the stronger atmospheric mixing for this planet limits their collision for either the nominal or reduced eddy profiles. Thus, for HD 209458 b, both the lower anticipated mass fluxes, as well as the high temperature and strong atmospheric mixing, limit the abundance of photochemical aerosols in its atmosphere. As the observational constraints for the thermal structure of HD 209458 b's upper atmosphere (above 1 mbar) are limited, we can inversely argue that the temperature in the upper atmosphere of this planet  must be hotter than on HD 189733 b, in order to explain the lack of significant hazes in the primary transit observations.

\begin{figure*}
\centering
\includegraphics[scale=0.45]{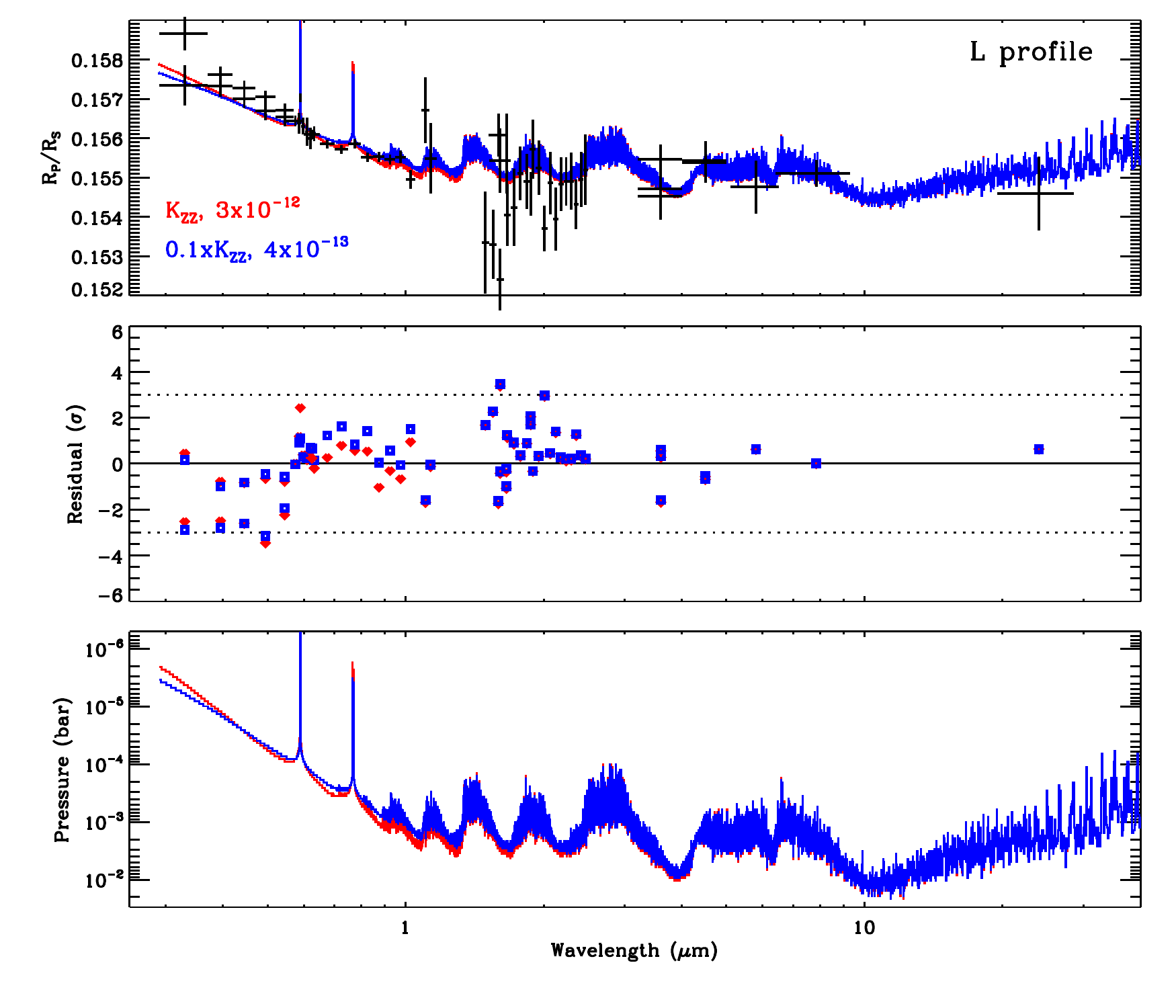}
\includegraphics[scale=0.45]{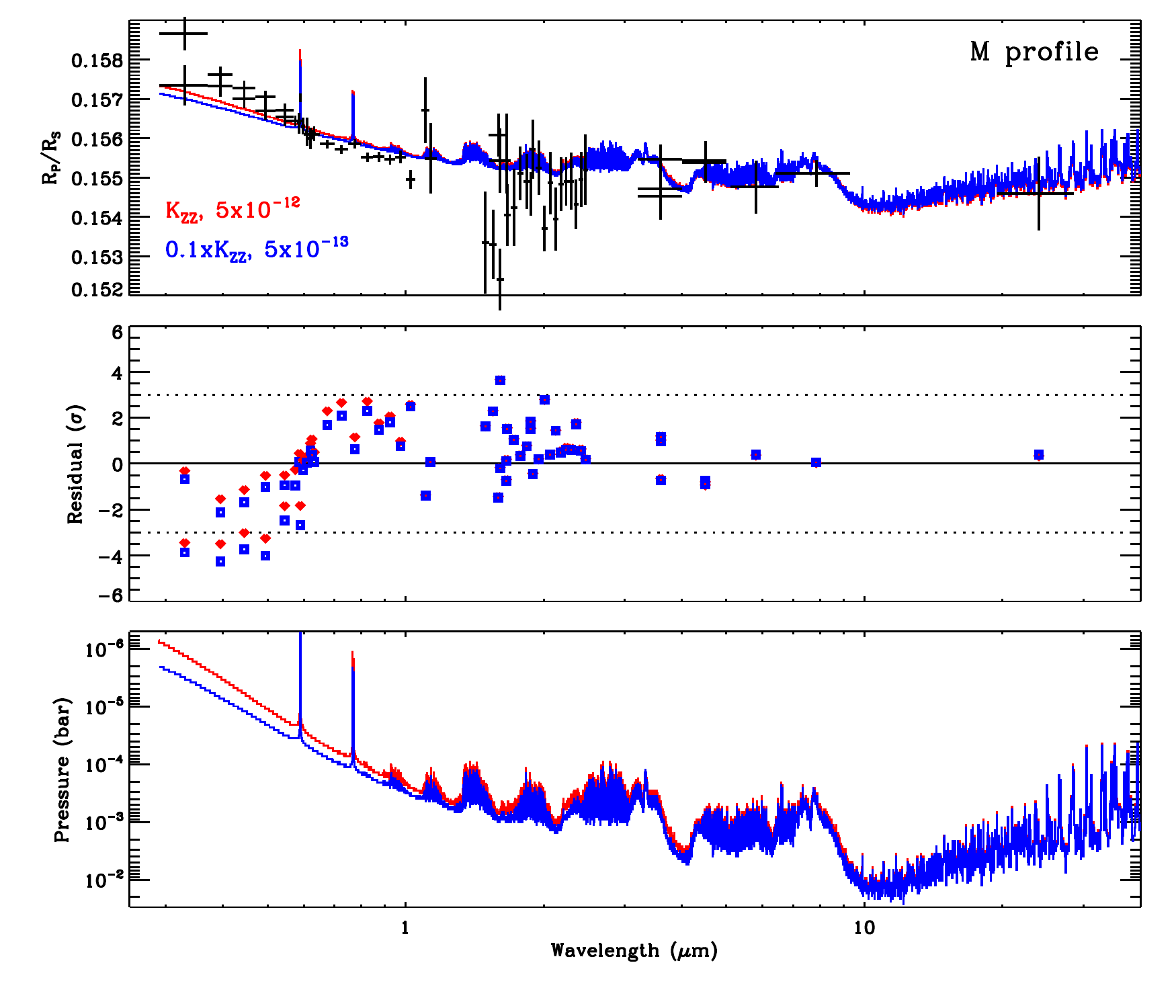}
\caption{Planetary transit radius for an aerosol mass flux of 5$\times$10$^{-13}$ g cm$^{-2}$s$^{-1}$ (upper panel). The pink and cyan lines correspond to the M and L temperature profiles, respectively. The red and blue squares present the model mapped on the resolution of the observations. The latter are taken from the compilation of \cite{Pont12}. The simulated spectra are within 3$\sigma$ of the observations in the whole spectrum (middle panel), although they appear systematically lower than the observed planetary size at UV wavelengths. The two temperature profiles provide a similar transit spectrum, although the reference pressure at the 8$\mu$m Spitzer/IRAC band is smaller for the M profile relative to the L profile (lower panel).}\label{transit2}
\end{figure*}

\subsection{Transit depth}

With the simulated aerosol distribution and the corresponding chemical composition from the photochemical model we can evaluate the simulated transit depth for each case. We calculate aerosol extinction at each altitude using Mie theory (spherical particles) and the refractive index for soot particles (details of this parameter are discussed below). For the gas components we consider Rayleigh scattering by H$_2$, collision induced absorption by H$_2$-H$_2$ and H$_2$-He pairs, extinction by Na and K, and absorption by H$_2$O, CO and CH$_4$. We calculate molecular lines using the high temperature line lists of Exomol \citep{Tennyson16} and Theorets \citep{Rey16}. For the planet size of HD 189733 b we use the reported $\rm R_P/R_S$ values from \cite{Pont12}, which are consistent with the latest review \citep{Sing16}. We calculate the transit depth by evaluating the attenuation of the stellar flux over the whole atmosphere \citep{Brown01}, and we use planet and star parameters from \cite{Boyajian14}.

There is a well-known degeneracy in the calculation of the planet size for gas giants due to the inability to know the atmospheric pressure at the observed planet size. In order to have a consistent picture among the different cases we evaluate, we normalise our simulated spectra to the reported planet size at the 8 $\mu$m band from Spitzer/IRAC observations. These observations are believed to be less affected by stellar limb effects and they demonstrate the smallest uncertainty from all near-IR observations \citep{Agol10}. To perform the normalisation we assume an initial reference pressure level for the reported $\rm R_P/R_S$, and with the corresponding altitude grid for the assumed temperature profile (calculated hydrostatically), we calculate the transit depth and the corresponding pressure probed at each wavelength. We then evaluate the average pressure in the 8 $\mu$m IRAC band taking into account the instrument response within the band \citep{Hora08}, and repeat the transit calculation assuming that the reported planet size at the 8 $\mu$m band corresponds to this average pressure. Thus, all simulated spectra are identical at the 8 $\mu$m band, but the reference pressure for each case is different, reflecting the changes in the gaseous and aerosol opacities for each case.

\begin{table*}[!t]
\centering
\caption{Best solutions in terms of required aerosol mass flux for each temperature and eddy case considered. We also provide, $\epsilon$, the implied the aerosol production efficiency relative to the mass fluxes provided in Table~\ref{mflux}, Pref, the reference pressure at the 8 $\mu$m band, the reduced $\chi^2$, and the average particle radii and the probed pressures at 300 nm and 500 nm.}\label{results}
\begin{tabular}{cccccccccc}
\hline
\hline
Profile	& Eddy			& Mass Flux			& $\epsilon$	& Pref	& $\chi^2$& r$_{300}$ 	& P$_{300}$	& r$_{500}$	& P$_{500}$\\
		&(K$_{\rm ZZ}$)	& (g cm$^{-2}$s$^{-1}$) 	& ($\%$)		& (mbar) 	& 		& (nm)		& ($\mu$bar) 	&(nm)		&($\mu$bar)	\\
\hline
L		& 1				& 3(-12)				& 10.35		& 2.34	& 1.7		& 1.5			& 2.5			& 2.3			& 59 \\
		& 0.1				& 4(-13)				& 2.85		& 2.28	& 1.8		& 1.7			& 3.8			& 4.7			& 48 \\
		& 0.01			& 6(-14)				& 0.21		& 2.43	& 2.7		& 3.4 		& 14			& 14.4		& 81\\
\hline
M		& 1				& 5(-12)				& 5.15		& 1.06	& 2.5		& 1.4			& 0.9			& 1.8			& 11\\
		& 0.1				& 5(-13)				& 3.57		& 1.48	& 2.7		& 1.6 		& 2.3			& 2.7			& 21\\
		& 0.01			& 7(-14)				& 0.072		& 1.66	& 3.8		& 2.3			& 8.0			& 8.3			& 43 \\
\hline
\hline
\end{tabular}
\end{table*}

\begin{figure}[!b]
\centering
\includegraphics[scale=0.5]{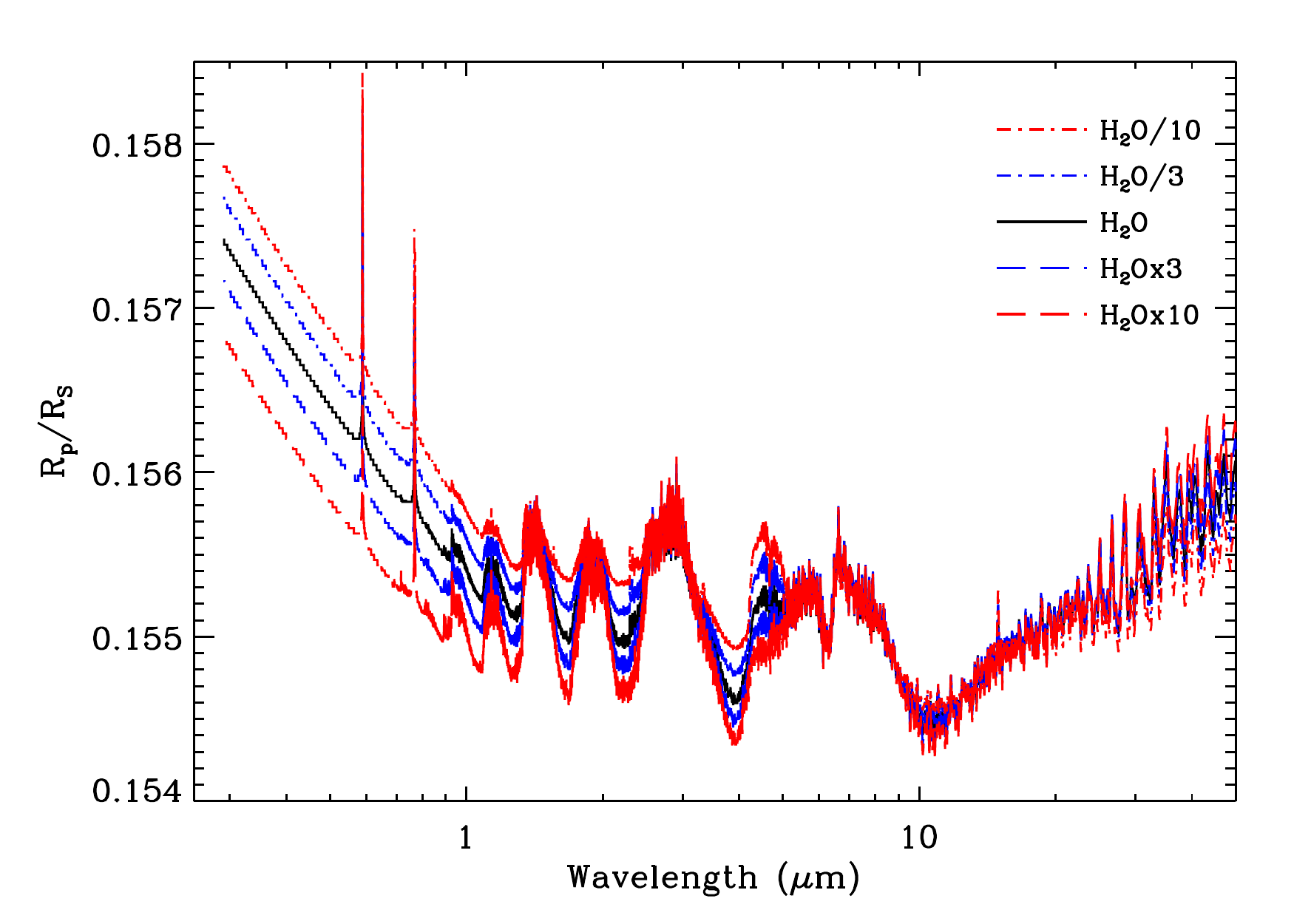}
\caption{Simulated planet sizes assuming different H$_2$O mixing ratios relative to solar abundances. Increasing the water abundance drops the UV planet size and vice versa decreasing the water abundance increases the contribution of aerosols.}\label{transit_H2O}
\end{figure}

We assume that the photochemical life times are much smaller than the characteristic times for the evolution of the microphysical aerosol population, i.e. the mass flux of aerosols is driven by the conditions at high illumination and once the particles are formed will be horizontally distributed. In contrast the gaseous composition will be modified locally due to the changes in the solar insolation. Thus, for the transit calculations we use chemical composition profiles that we calculate assuming limb geometry conditions for the photolysis rates (examples of limb profiles are presented further below). Temperature variations between the limbs and the disk can also affect the resulting transit size \citep{Line16}. However, we focus here on the differences incurred by the temperature profiles we consider, which demonstrate larger variations than those anticipated by the disk-limb variations. For example, the lower temperatures suggested by the M profile in the lower atmosphere allow for a higher abundance of CH$_4$ to survive in the upper atmosphere. This results in clear signatures in the simulated transit spectra near the strong $\nu_3$ (3.3 $\mu$m) and $\nu_4$ (7.7 $\mu$m) methane bands, which are missing in the warmer L profile case that is dominated by $\rm H_2O$ absorption. Current observations do not resolve these spectral regions, but future measurements with JWST should be able to identify such features, if present. Nevertheless, the overall picture of the transit spectra for these two extreme profiles is rather similar and only the reference pressure level changes between the two cases. 

\begin{figure*}[!t]
\centering
\includegraphics[scale=0.45]{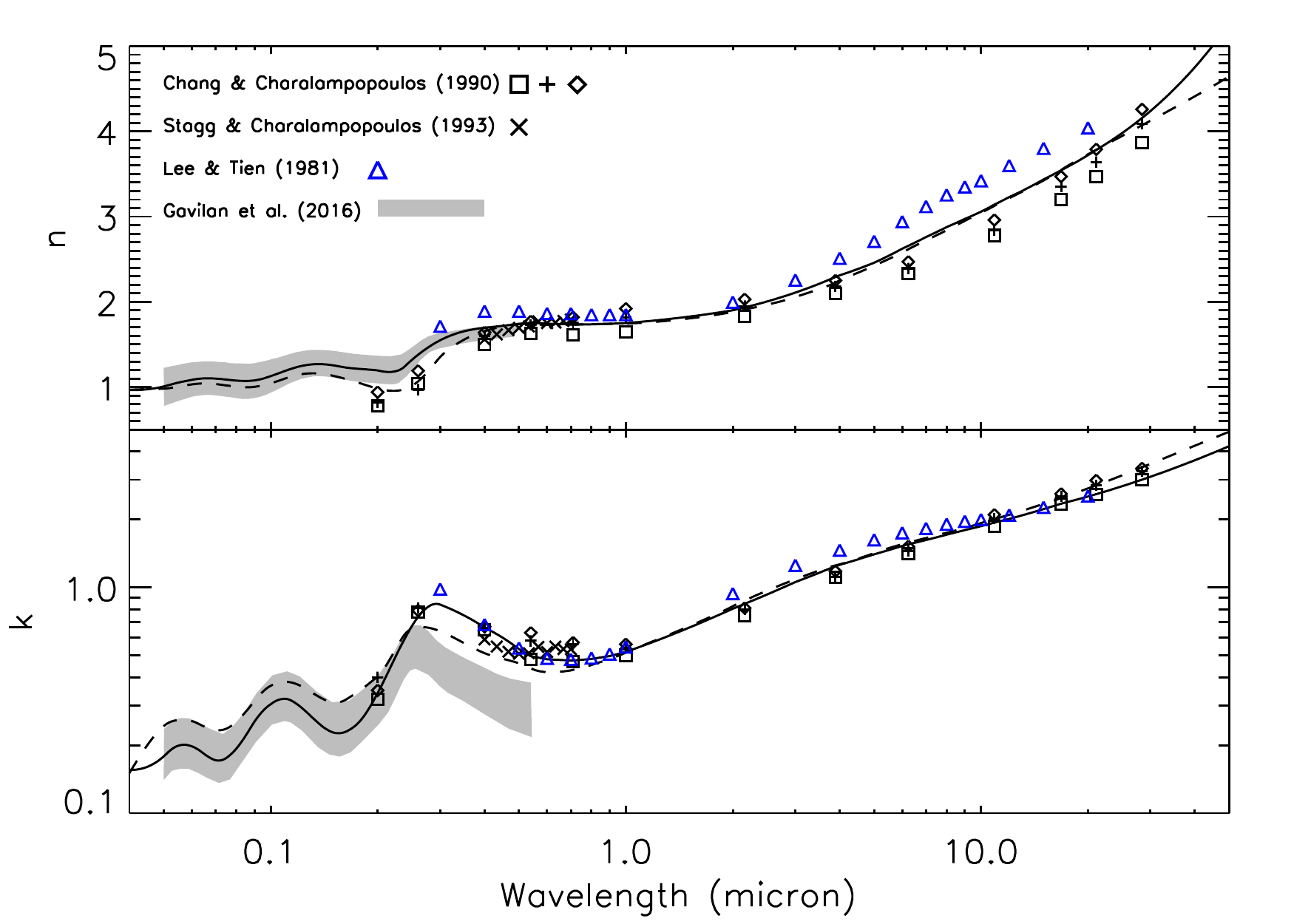}
\includegraphics[scale=0.45]{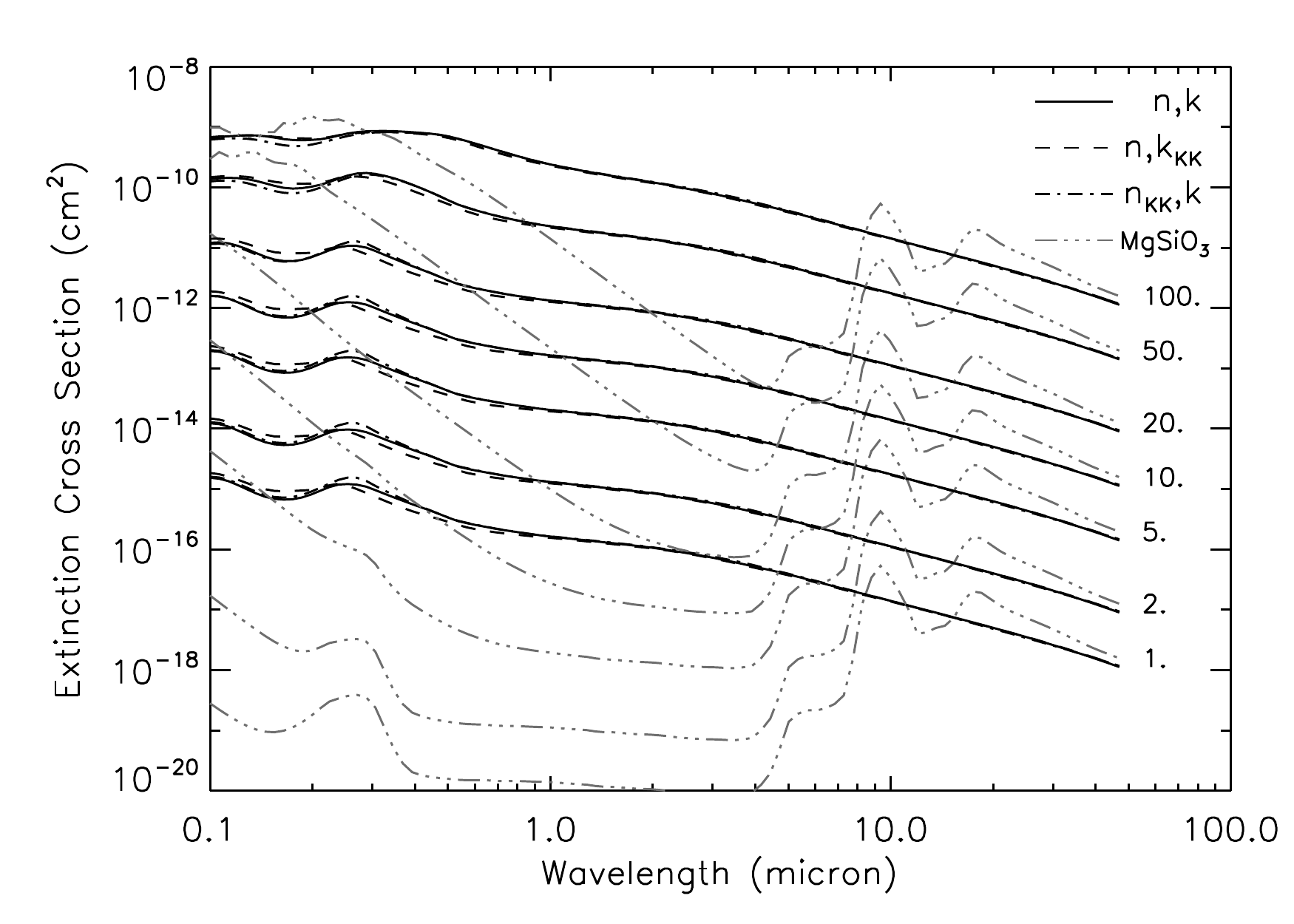}
\caption{Left: Compilation of refractive index measurements of soot particles from different laboratory studies (symbols). The solid lines present average spectra for the real and imaginary parts, and dashed lines calculated components based on the Krammers-Kroning dispersion formulae. Right: Extinction cross section of soot composition particles at different radii (in nm). Different curves correspond to different evaluations of the refractive index (see text).  The gray lines present the equivalent cross section assuming silicate composition (MgSiO$_3$).}\label{Soot_ri}
\end{figure*}

The impact of the simulated aerosol distribution on the planet size is mainly affecting the UV and visible part of the spectrum, as anticipated by the small particle radius (Fig.~\ref{transit1}). The resulting $\rm R_P/R_S$ spectra demonstrate how the planet size increases at short wavelengths as the assumed aerosol mass flux increases. The simulated spectra clearly show the inverse relationship between the aerosol mass flux and the atmospheric eddy mixing, with lower mixing requiring smaller aerosol mass fluxes to match the observations. For the nominal eddy profile, mass fluxes between 10$^{-12}$ and 10$^{-11}$ g cm$^{-2}$s$^{-1}$ bracket the observations for both temperature profiles (L $\&$ M), while for weaker atmospheric mixing the required mass flux is smaller and a flux of 10$^{-12}$ g cm$^{-2}$s$^{-1}$ provides an upper bound. This effect results from the larger particle radius accessible for the low atmospheric mixing efficiencies, which allows for an increase of the particle cross section that outbalances the drop of the corresponding particle density (see Fig.~\ref{micro}). However, the particle radius modifies the slope of the particle extinction at short wavelengths, which defines the shape of the simulated spectra. Small particles lead to steep slopes that progressively become shallower as the particle radius increases. This trend is obvious in the simulated spectra, particularly for the lowest eddy case (0.01$\times$K$_{\rm ZZ}$) that provides the largest particle radii ($\sim$ 10 nm at the probed pressures). In this case the simulated transit spectrum has a much shallower slope at short wavelengths compared to the nominal eddy case for which the particle radius remains small ($\sim$1-2 nm at the probed pressures).

If we attempt to fit the observation with the current assumptions, we find that for the lowest eddy case (0.01$\times$K$_{\rm ZZ}$) a mass flux of the order of 6-7$\times$10$^{-14}$ g cm$^{-2}$s$^{-1}$ is required (Table~\ref{results}). Such a mass flux implies a photochemical aerosol production efficiency of $\sim$0.1-0.2$\%$ relative to the mass fluxes involved into soot formation (Table~\ref{mflux}). The other atmospheric mixing cases imply a progressively larger efficiency increasing from 3$\%$ to 10$\%$ as the eddy is increased from 0.1$\times$K$_{\rm ZZ}$ to the nominal case. The required mass flux for the nominal eddy case (3$\times$10$^{-12}$ g cm$^{-2}$s$^{-1}$) is higher than the 0.1$\times$K$_{\rm ZZ}$ case, by a factor of $\sim$10 for both temperature profiles (Table~\ref{results}). Moreover, the reference pressures derived for the L profile are systematically higher than for the M profile, which occurs due to the increased contribution of CH$_4$ for the latter case that increases the total atmospheric opacity.

The simulated spectra for the optimal solutions (in terms of their reduced $\chi^2$, see Table~\ref{results}) demonstrate the increasing planet size towards short wavelengths, while they indicate that solutions under the L temperature profile provide a slope that is closer to the observations relative to the M temperature profile, for both assumed eddy mixing cases (Fig.~\ref{transit2}). The residuals between model and observations are within $\sim$3$\sigma$ of the measurement uncertainty for the L case, while for the M case the model difference from the high-transit depth cases at short wavelengths approaches  $\sim$4$\sigma$. The higher temperature at the probed pressures for the L case, imposes a larger atmospheric scale height relative to the M case that increases the slope of the spectrum. Our results indicate that the average particle radius roughly doubles going from $\lambda$=300 nm to $\lambda$=500 nm, while the probed pressures respectively increase from a few $\mu$bar to $\sim$50$\mu$bar (Table~\ref{results}). At those pressures we are near the aerosol production altitude, thus the assumptions made for the shape and location of the aerosol production profile can potentially affect the resulting average particle size and density, therefore the simulated spectra at short wavelengths.
 
Another aspect that can affect the transit size at short wavelengths is the elemental composition, which we assume to be solar in the chemical composition simulation. Since the spectra are normalised to a common reference point (the 8 $\mu$m IRAC band), modifying the mixing ratio of H$_2$O while keeping all other components the same would affect the planet size at short wavelengths relative to the planet size in the visible-near-IR, which is dominantly affected by water absorption; a higher (lower) water abundance would cause light from the star to be attenuated at a higher (lower) altitude, therefore bring the planet size in the IR closer to (further away from) the planet size observed in the UV (see Fig.~\ref{transit_H2O}). Thus, given the current uncertainty on the elemental composition of HD 189733 b's atmosphere and that of the observed transit depth, it is pointless to seek for a better fit to the observations.

\section{Haze interaction with the atmosphere}

We now focus our attention on the possible implications that photochemical aerosols could have on the atmospheric structure and composition, and on how the particles interact with the gas phase background. To perform this investigation we need to know how the formed particles will interact with the radiation field. For this task we used a compilation of laboratory investigation for the refractive index of soot particles (see Fig.~\ref{Soot_ri}). These are based on the measurements of \cite{Chang90} and \cite{Lee81} in the visible and IR, and the \cite{Gavilan16} measurements in the UV. To check the consistency between the real and imaginary parts of the refractive index we used the Krammers-Kroning dispersion relations. Thus, we derived average spectra of n and k from the available measurements (solid lines in Fig.~\ref{Soot_ri}), from which we calculated the corresponding k and n through the dispersion formulae (dashed lines). The results suggest that the average n and k spectra are consistent with their corresponding measurements within the uncertainty of the derived laboratory values, which also rely on the dispersion formulae for their evaluations. Moreover, the observed differences translate to minor changes in the resulting extinction cross sections of the aerosol particles (see Fig.~\ref{Soot_ri}). Therefore, in the following we used the average n and k spectra (solid black lines). 

\begin{figure}[!t]
\centering
\includegraphics[scale=0.5]{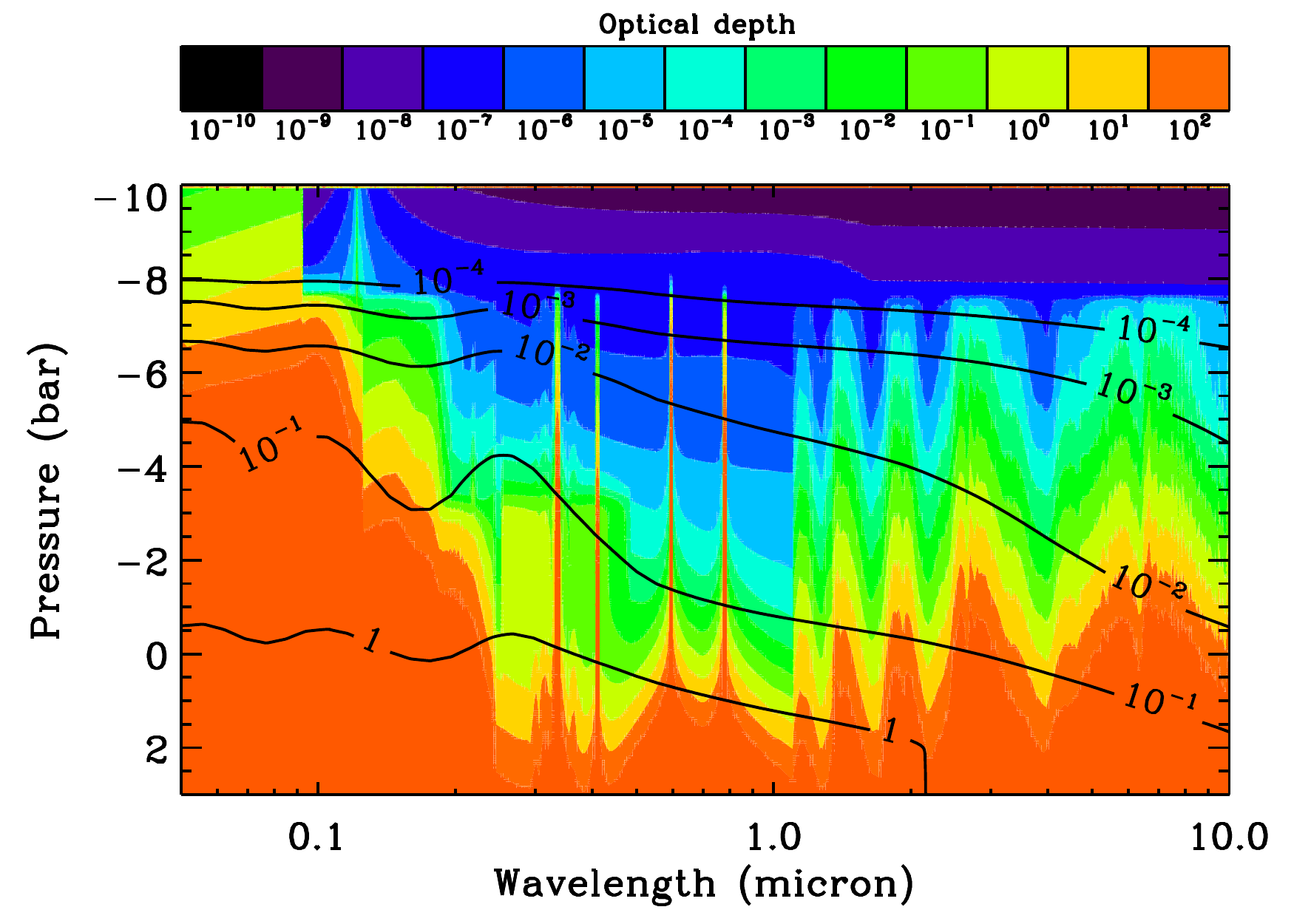} %Gas_ext_new.eps}  % M profile, Kzz, 5.E-12 
											% M profile, Kzz, 1.E-13
											% M profile, 0.1xKzz, 5.E-12
											% L profile, Kzz, 3.E-12
\caption{Wavelength depended atmospheric opacity at different locations in the atmosphere of HD 189733 b {\color{\clr} for the M profile and under nominal eddy conditions}. The background coloured contours present the contribution of gaseous components, while the foreground black line contours correspond to the aerosol {\color{\clr}extinction} contribution.}\label{opacity}
\end{figure}

\subsection{Radiation field $\&$ chemistry}

\begin{figure}[!t]
\centering
\includegraphics[scale=0.5]{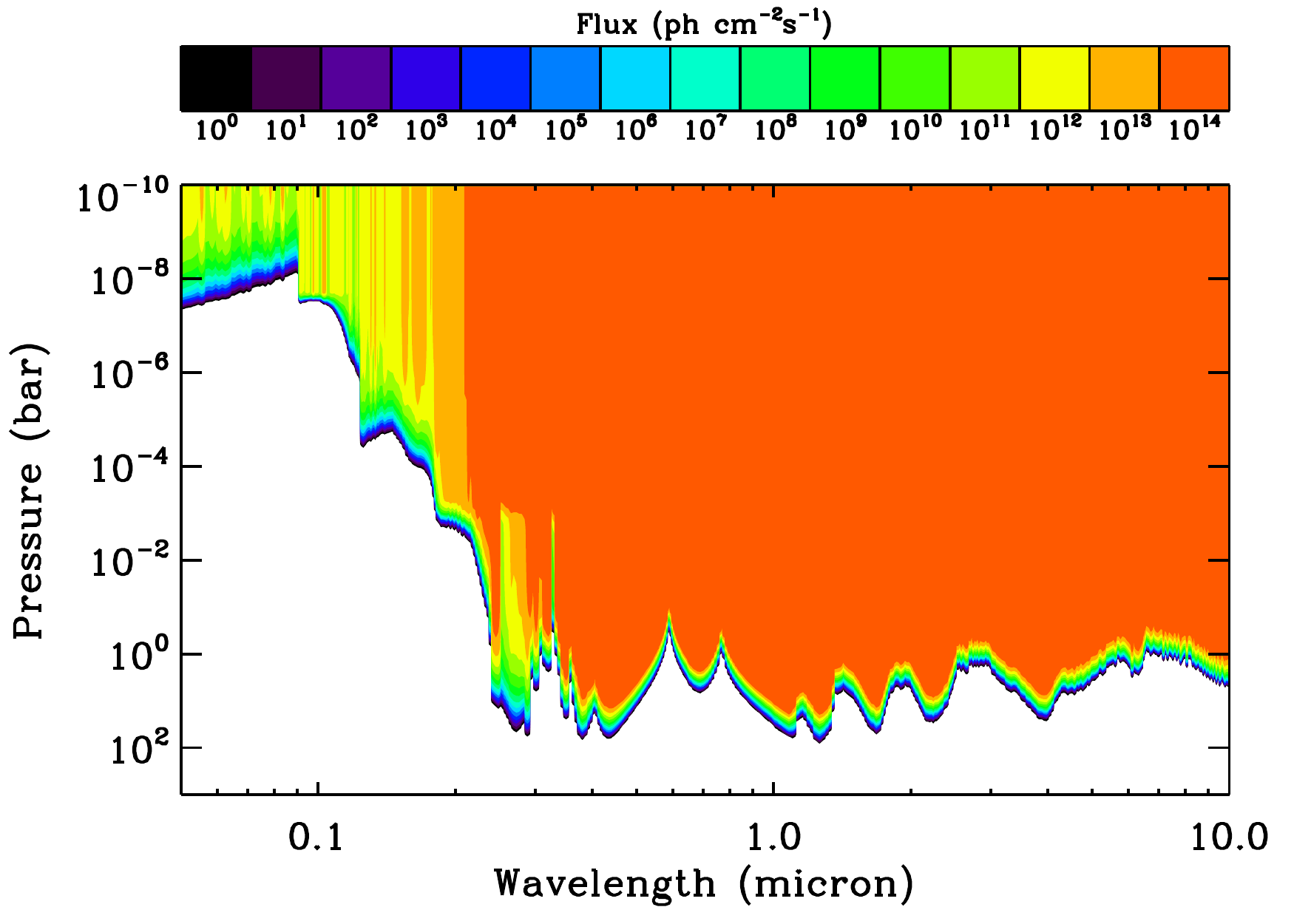}
\caption{Contour plot of actinic fluxes vs wavelength in the atmosphere of HD 189733 b for disk average simulation (M profile).}\label{flux}
\end{figure}

With the simulated aerosol particle distribution and cross sections we can subsequently evaluate their contribution to the atmospheric opacity. Taking as an example the case of the disk average simulations for the M profile (see Fig.~\ref{opacity}) we can compare the total extinction optical depth due to the gaseous components with the corresponding opacity from the photochemical aerosols. The total aerosol optical depth is small compared to the total gaseous contribution in the deep atmosphere, while at short wavelengths ($\lambda$$<$1000 $\rm\AA$), absorption by hydrogen dominates the atmospheric opacity and photons at those wavelengths are absorbed at altitudes above the region of the aerosol contribution (Fig.~\ref{flux}). However, at longer wavelengths the vertical distribution of aerosol particles provides a local opacity in the middle atmosphere (p $<$ 10 mbar) that is larger than or  comparable to the gaseous contribution. The importance of aerosol contribution varies across the wavelength range due to the strong variation of the gaseous opacity mainly from the alkali lines in the visible and from H$_2$O absorption bands in the near IR. Nevertheless, the presence of the aerosols will affect both the photolysis rates of different species  as well as the local atmospheric heating.

\begin{figure*}
\centering
\includegraphics[scale=0.5]{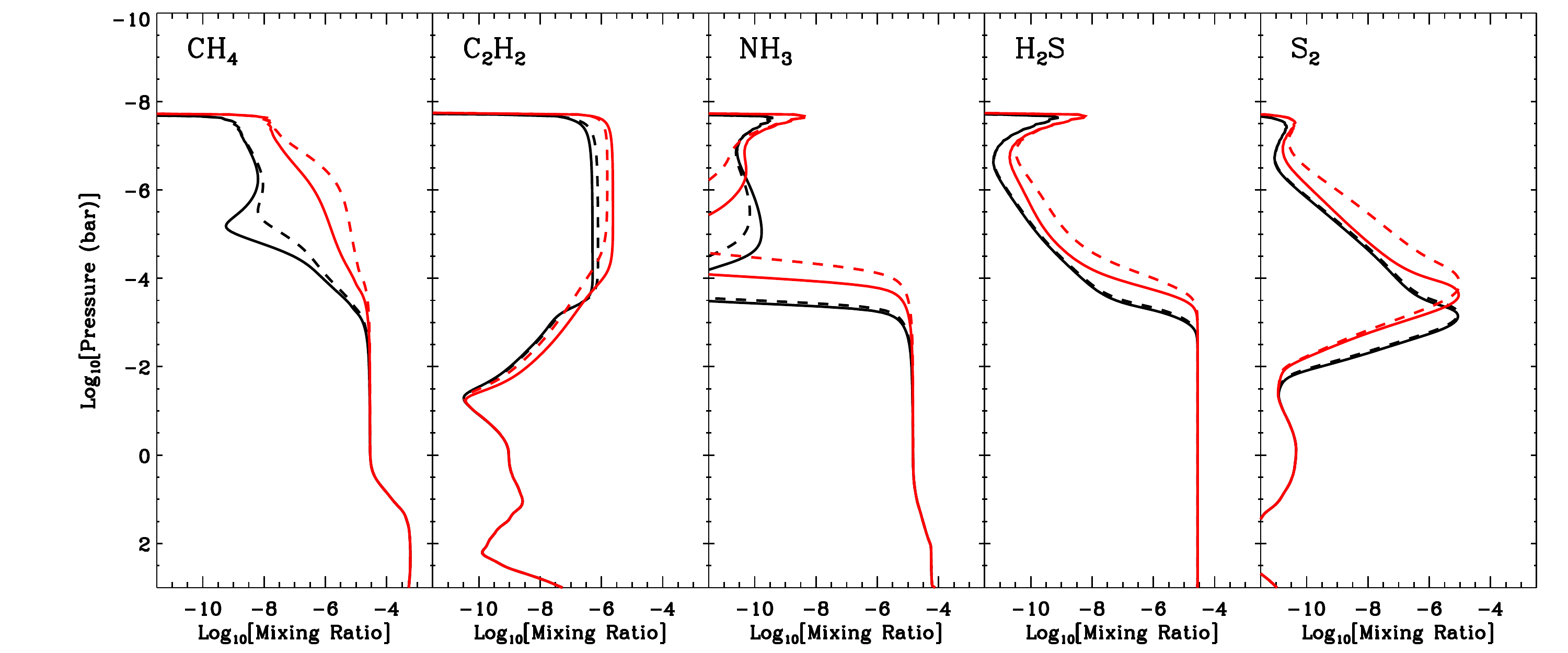}
\includegraphics[scale=0.5]{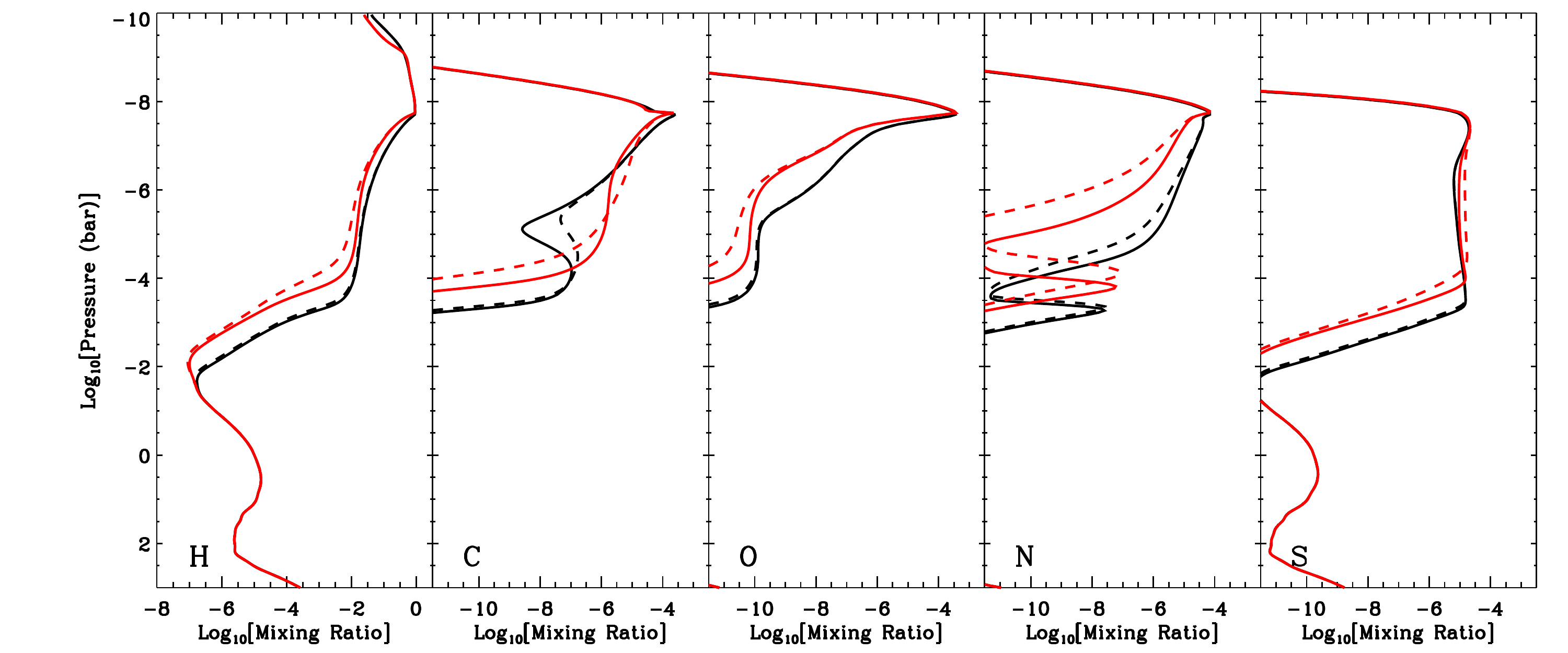}
\includegraphics[scale=0.5]{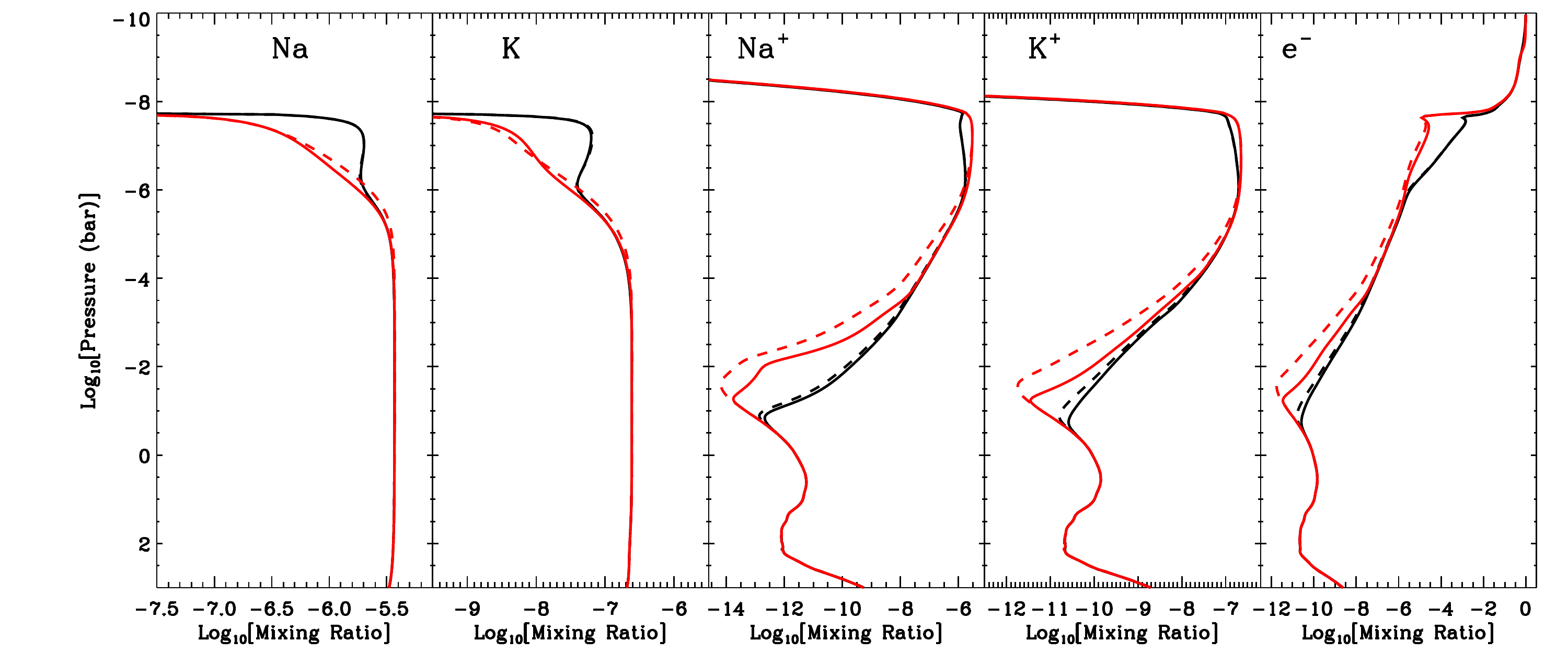}
\caption{Mixing ratio profiles for different species in the atmosphere of HD 189733 b assuming the M profile. The black and red lines correspond to disk average and limb geometry conditions, while the dashed and solid curves present the results with and without the impact of aerosols on the radiation field.}\label{aer_chem}
\end{figure*}

% Composition 
Photons at wavelengths between 100 and 300 nm partake in photochemical processes of molecules larger than H$_2$, therefore the modifications imposed by the aerosols affect the gaseous abundances. Our chemistry model results (Fig.~\ref{aer_chem}) demonstrate this effect when the simulated aerosol opacity is included in the evaluation of the atmospheric photolysis rates. These are reduced relative to our previous calculations of a clear atmosphere, since a fraction of the available photons at each atmospheric location is absorbed (or partially scattered) by the aerosols. Thus, the abundances of main species such as CH$_4$, NH$_3$, and H$_2$S that have cross section in the wavelength range of aerosol influence are increased. Secondary products (e.g. C$_2$H$_2$, S$_2$) that have photolysis cross section that extend further in the FUV, are also directly affected by the aerosol opacity, but also indirectly through changes the aerosols impose on the chemical abundances of intermediate species leading to their formation.

For the simulated aerosol opacity required to match the transit observations, the impact on the disk average composition results is small and pertains mainly to the methane profile. However, the longer path-lengths of the limb geometry increase the contribution of the aerosol opacity, therefore the impact on the gaseous composition is larger for this case (Fig.~\ref{aer_chem}).

The simulated changes in atmospheric composition from the inclusion of the aerosol opacity mainly concern the upper part of HD 189733 b's atmosphere. Therefore, they do not affect the interpretation of broadband observations from space telescopes, which are sensitive to pressures greater than 1-10 $\mu$bar, depending on the thermal structure assumed (see Fig.~\ref{transit2}). However, for high-resolution ground based observations, as for example the Na transit observations \citep{Wyttenbach15}, the modifications of the Na profile imposed by the aerosols (Fig.~\ref{aer_chem}) could have an observable signature, given the sensitivity of the transit at the line cores to the Na abundance in the upper atmosphere. We will investigate this aspect in more detail in a forthcoming study.

\subsection{Atmospheric heating}

\begin{figure}[!t]
\centering
\includegraphics[scale=0.5]{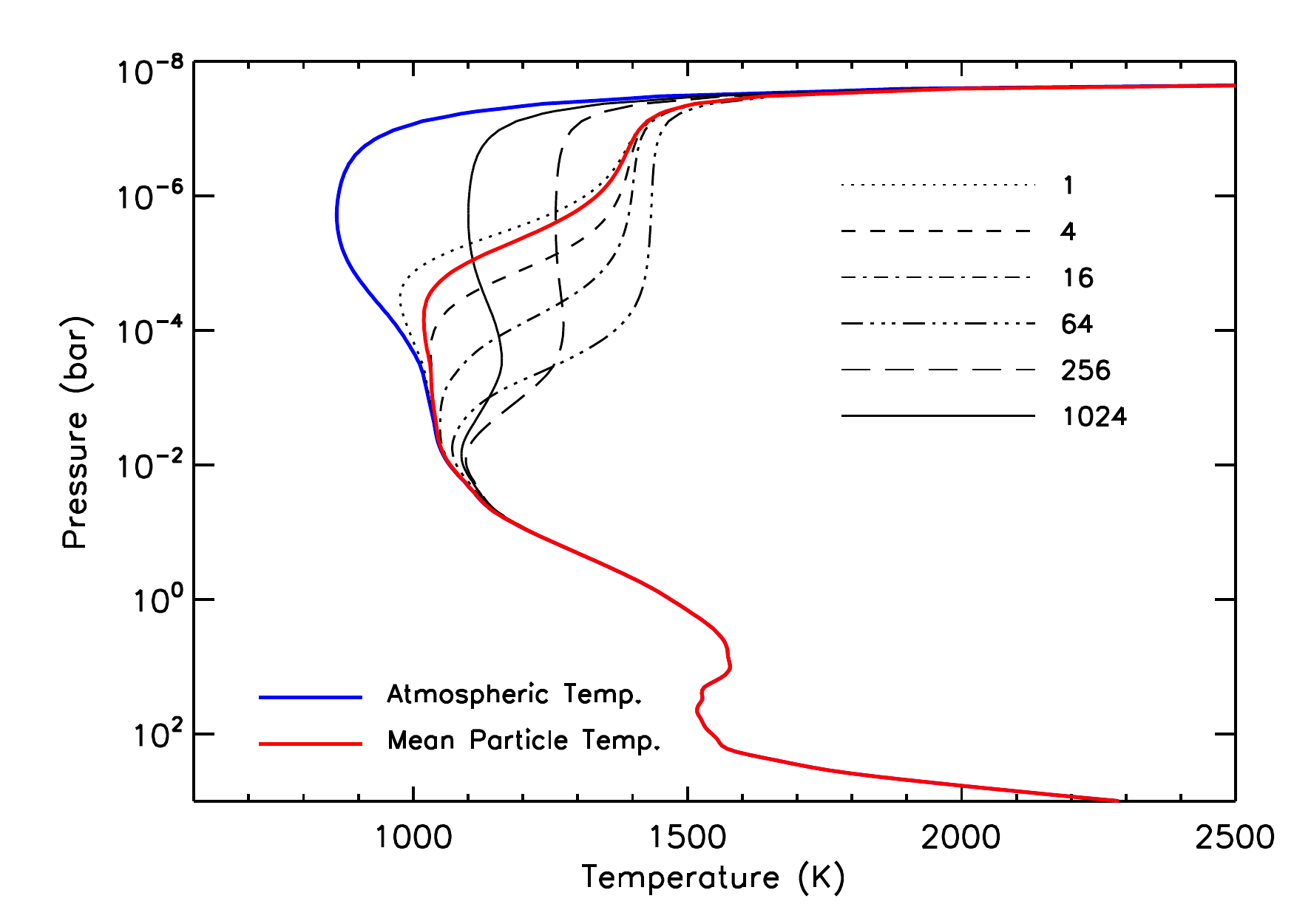}
\includegraphics[scale=0.5]{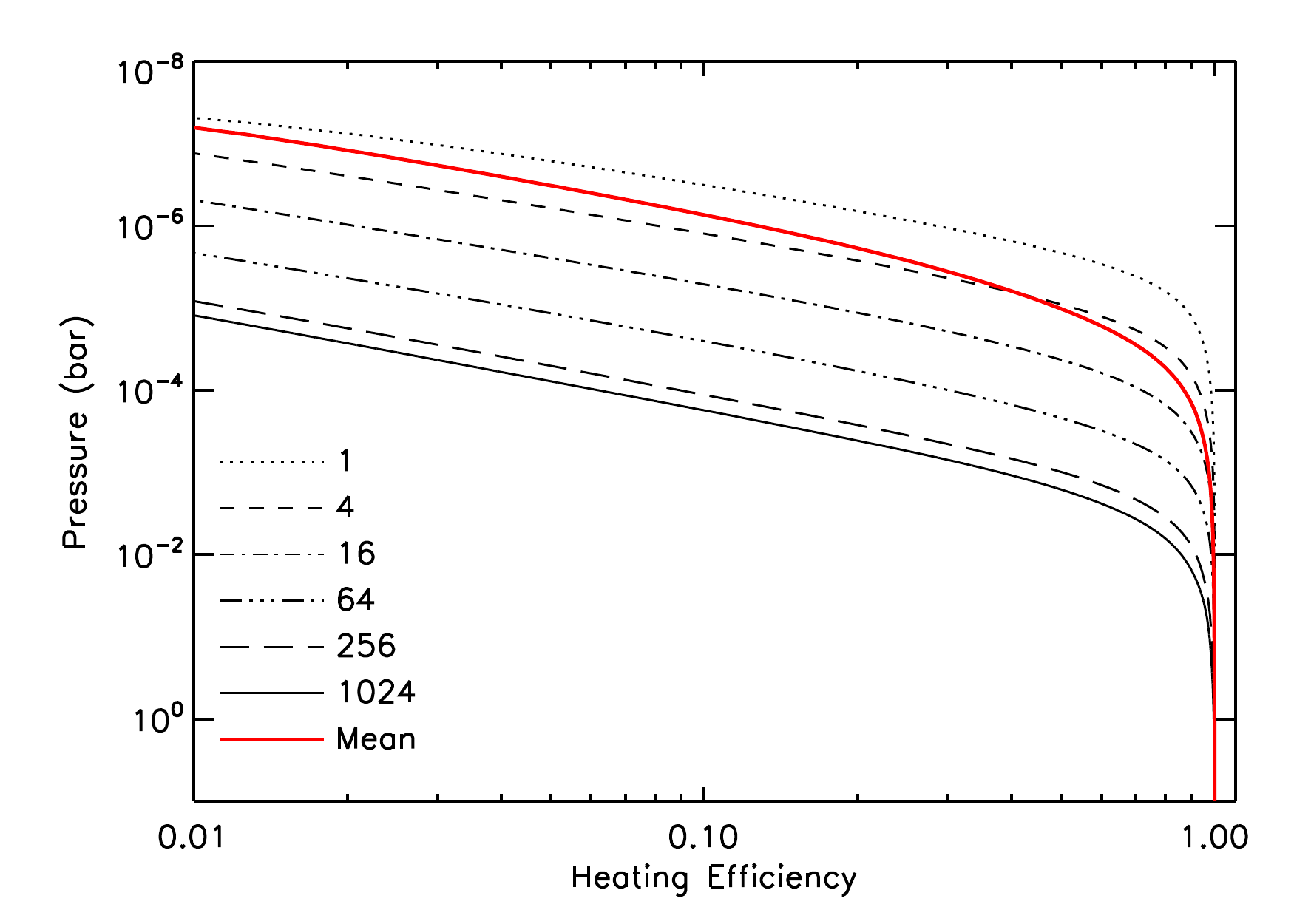}
\caption{Top: Particle temperatures (black lines) for different radii (in nm) relative to the atmospheric background temperature (blue). The red line presents the average particle temperature based on the simulated size distribution. Bottom: Particle heating efficiencies (black lines) at different locations in the atmosphere of HD 189733 b, and the corresponding average profile (red line).}\label{efficiency}
\end{figure}

\begin{figure}[!t]
\centering
\includegraphics[scale=0.5]{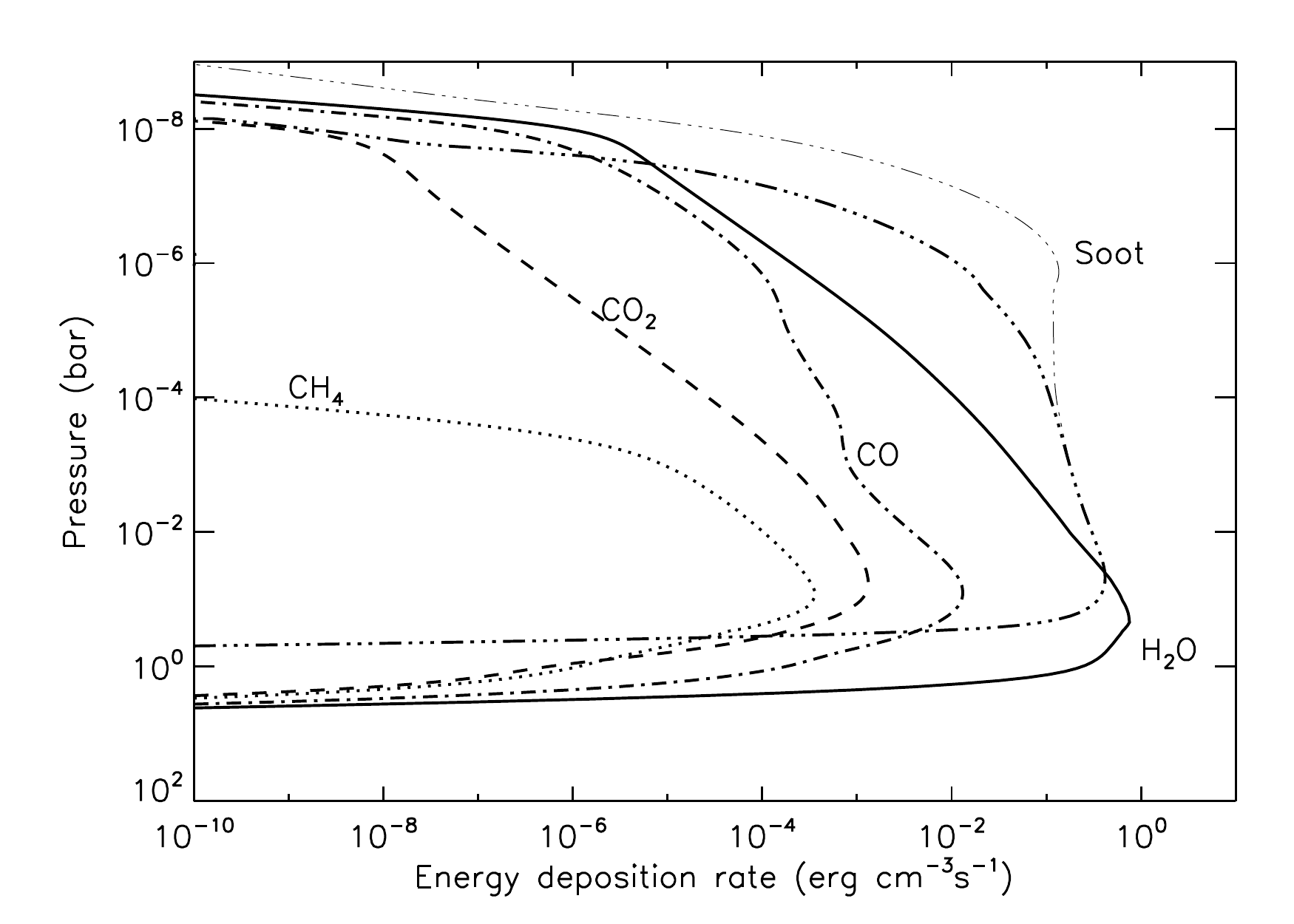}
\includegraphics[scale=0.5]{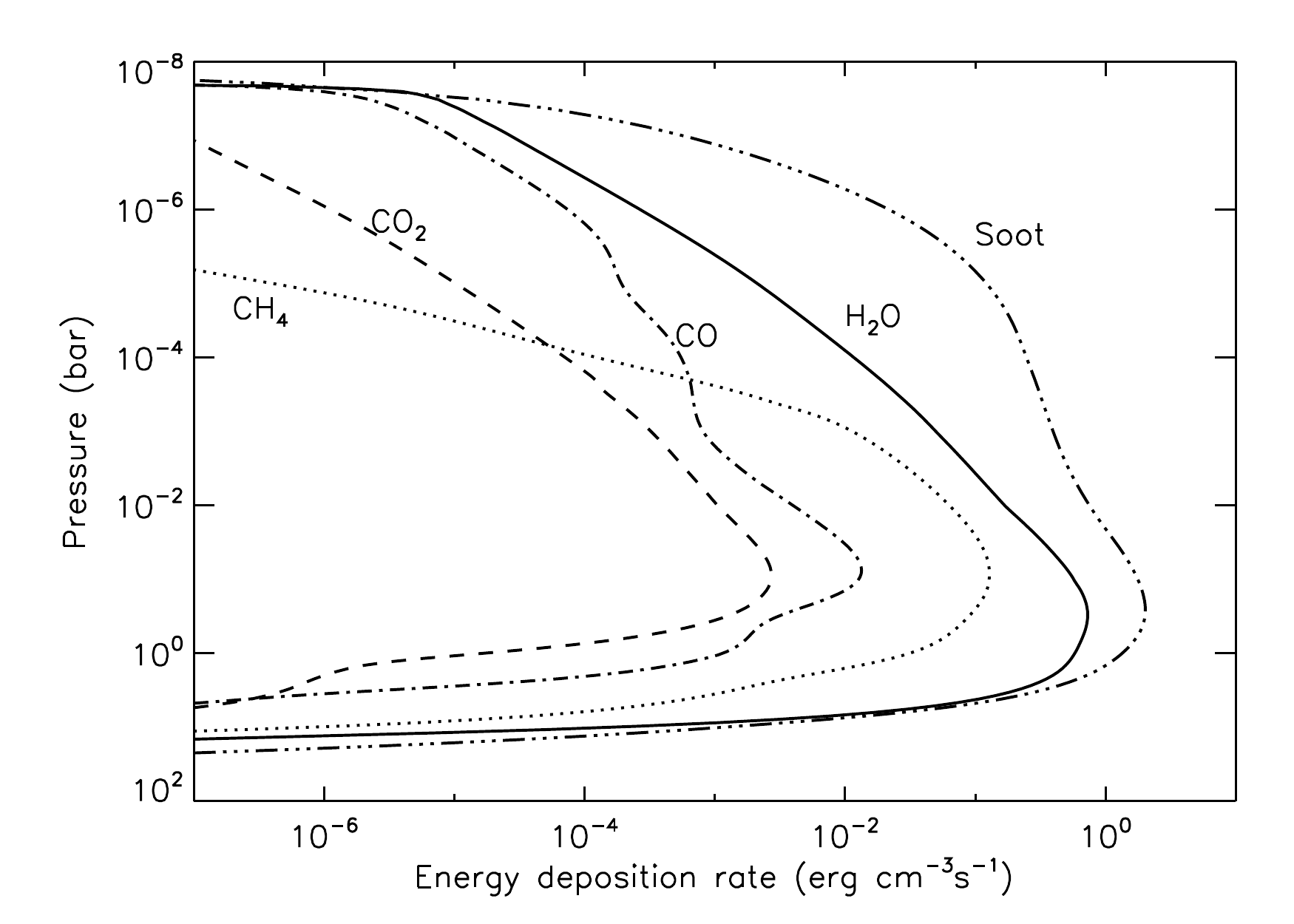}
\caption{Energy deposition rates from the absorption of visible and near-IR radiation by the major gaseous components and the aerosols. The top panel presents results for the L profile and bottom panel for the M profile (see text). The thick and thin lines for the aerosol contribution in the L case present the resulting atmospheric heating with and with out the inclusion of the particle heating heating efficiency.}\label{heating}
\end{figure}

Apart from their impact on the photochemistry, aerosols can have a major influence on the atmospheric thermal structure as indicated by their opacity at visible wavelengths, relative to the contribution of gases. In order to evaluate their impact we calculated atmospheric heating rates from the major gaseous absorbers (H$_2$O, CO, CO$_2$, and CH$_4$), as well as heating from aerosol absorption (Fig.~\ref{heating}). For the gaseous heating we considered absorption of the stellar radiation from the molecular bands taking into account the attenuation of the incoming radiation from all atmospheric components including extinction at the Na and K resonant lines and Rayleigh scattering by atomic and molecular hydrogen. 

For the aerosol heating rate we can not assume that all absorbed radiation is transferred to the atmosphere as heat, because the aerosol population occupies a low density part of the atmosphere. At these conditions, the particles are heated through absorption of stellar radiation but if collisions are not rapid in order to equilibrate their temperature with the background atmosphere, part of the absorbed energy will be emitted as IR radiation that will partially escape to space. Therefore the atmospheric heating efficiency due to the aerosols will decrease with lower atmospheric density. To evaluate this parameter we calculated the surface temperature of each particle in our size distribution at each atmospheric location, by equilibrating the local heating and cooling rates. Heating is due to the absorption of the incoming stellar radiation, while cooling occurs due to thermal emission at the particle's surface temperature (considered to be a black body with an emission efficiency equal to the absorption efficiency at each wavelength) and energy transfer through collisions with the bulk atmospheric molecules. For the last term we considered the thermal conductivity of a H/H$_2$ gas mixture, a complete thermal accommodation ($\alpha_{\rm T}$=1), and the transition from the continuous to the free-molecular regime in our evaluation of the molecular collision rates with the aerosol particles \citep[see][for more details]{Lavvas11c}. Our calculations indicate that the particle temperature starts to diverge from the atmospheric background temperature above 10 mbar for the largest particles resulting from our simulations, thus, the heating efficiency of the particles decreases at lower pressures (Fig.~\ref{efficiency}). A small part of the emitted IR radiation may be re-absorbed by the atmosphere, thus our calculations provide a minimum estimate for the impact of the aerosols on the atmospheric heating rate. 

The simulated energy deposition rates demonstrate that H$_2$O provides the major heating in the lower atmosphere near the 1 bar level (Fig.~\ref{heating}). However, at lower pressures (below $\sim$10 mbar) the photochemical aerosols dominate over all atmospheric heating contributions. This conclusion applies to both temperature profiles considered, although for the lower temperature M profile, for which aerosols can survive to deeper pressures and reach higher particle sizes, their corresponding atmospheric heating is higher and dominates over the H$_2$O contribution in the whole atmosphere. The relative contributions of the gaseous opacities also vary among the two atmospheric profiles considered, with more characteristic the heating by methane that is second after water vapor for the M profile, but practically insignificant for the L profile. This is a direct reflection of the different methane abundances allowed for each profile, as described above. Irrespective of the atmospheric structure assumed, we see that the presence of photochemical aerosols could significantly affect the atmospheric thermal structure and therefore contribute in merging the theoretical and retrieved temperature profiles. Thus their impact should be considered in future studies of EGP thermal structure.

\begin{figure*}[!t]
\centering
\includegraphics[scale=0.45]{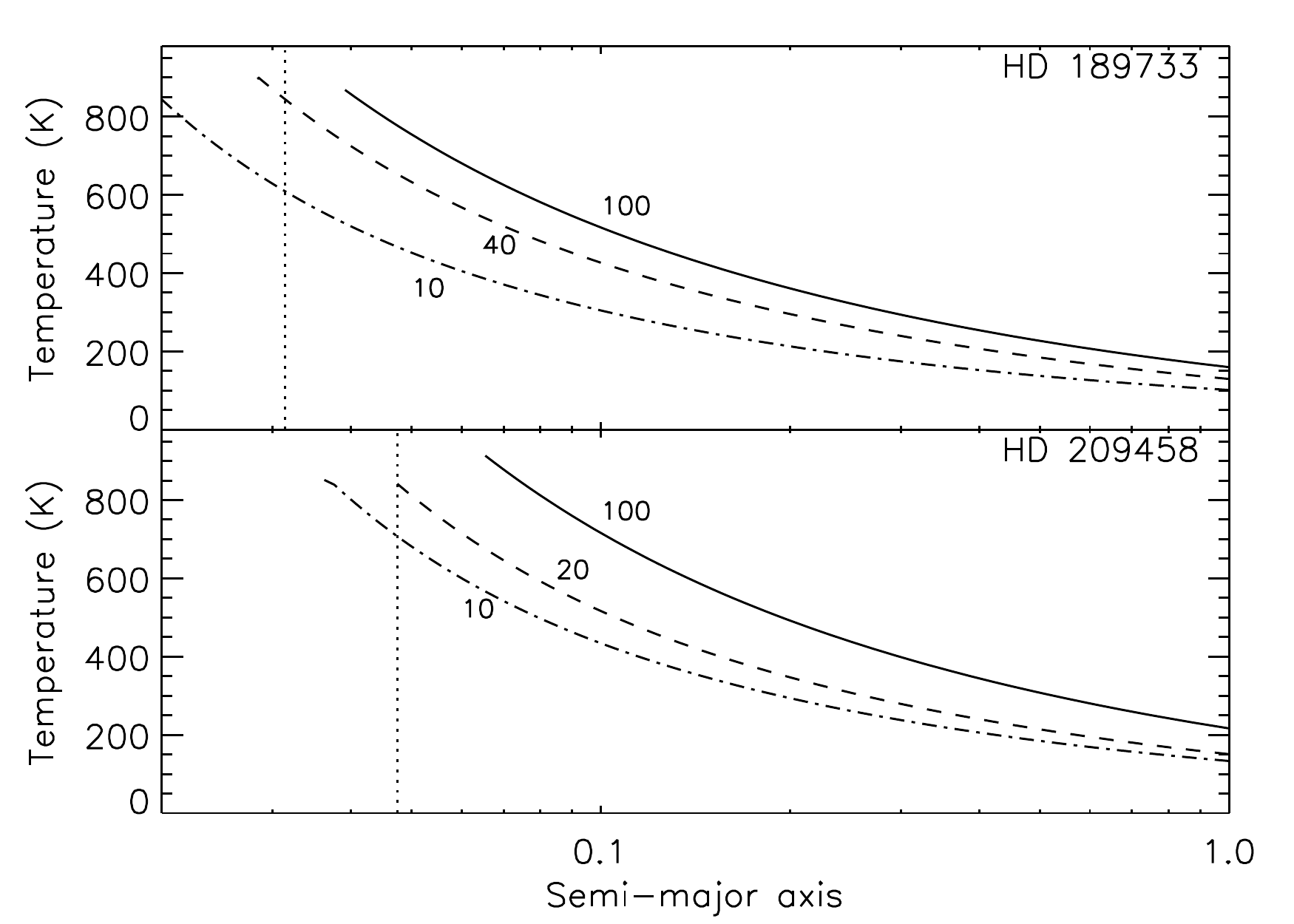}
\includegraphics[scale=0.45]{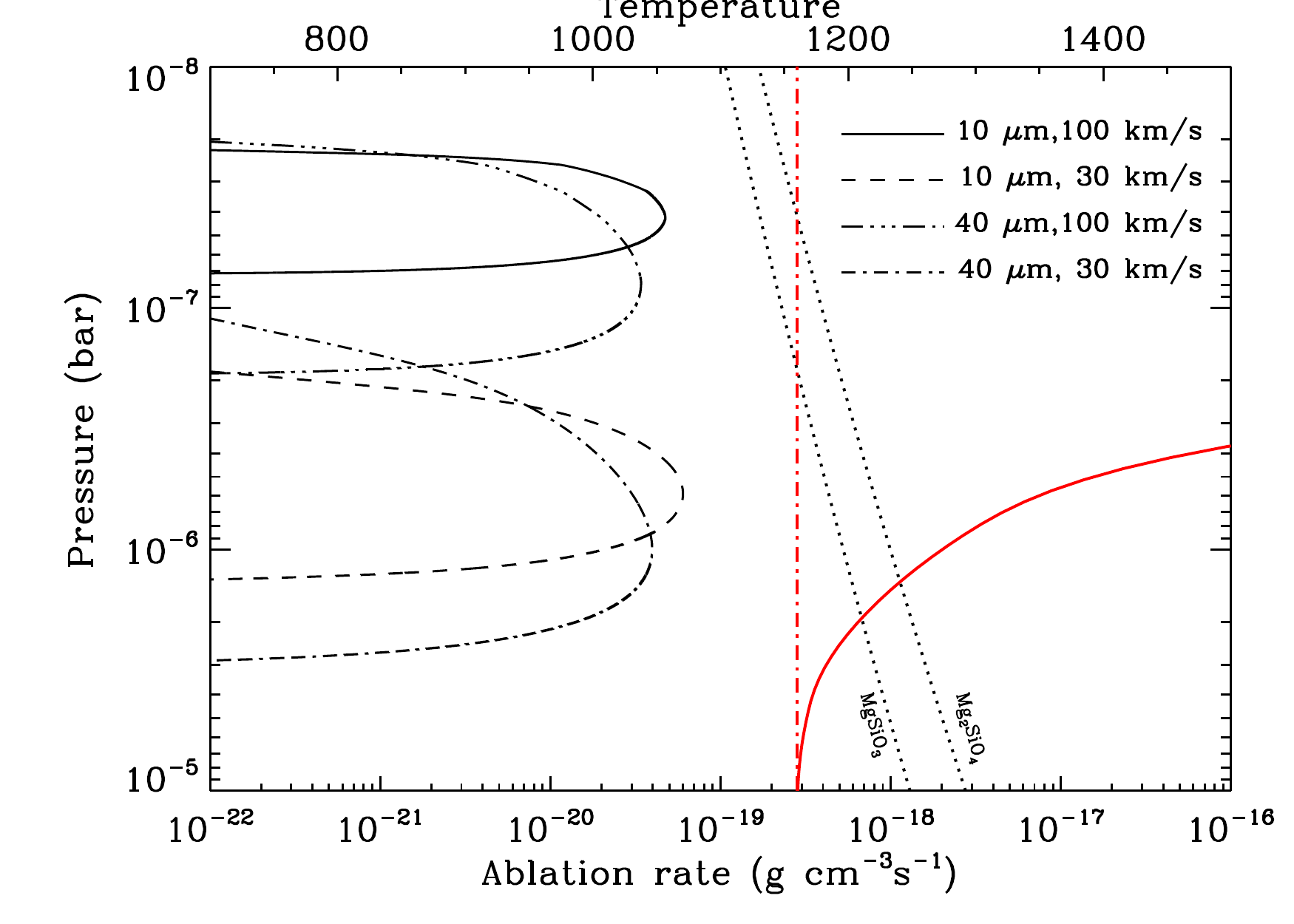}
\caption{Left: Variation of meteoroid surface temperature and radius at different orbital distances around HD 189733  and HD 209458. Each curve presents a different initial radius as designated (in $\mu$m) and meteoroids are considered to have a silicate composition. Right: Meteoroid ablation rate in the atmosphere of HD 189733 b for different meteoroid radii and impact velocities (black curves).  The incoming meteoroid mass flux is assumed to be 10$^{-12}$ g cm$^{-2}$s$^{-1}$.The red lines present the thermal structure for the L and LK profiles in the region of the atmosphere where meteoroids ablate. Dotted lines present the condensation curves for silicate (see Fig.~\ref{pT}).}\label{meteoroids}
\end{figure*}

\section{External sources}

So far we considered processes within the atmosphere that could affect the formation of photochemical aerosols. However, the contribution of external sources such as meteoroids should be evaluated. Meteoroids in the solar system have a composition reflecting their formation region, that mainly includes a metallic core dominated by iron and silicate minerals covered by ices \citep{Jarosewich90}. At the orbital distances of hot-Jupiters ($\lesssim$0.1 AU) the surface temperature of the incoming particles will be high and the icy coating of the meteoroids will evaporate. Thus, if we consider only metallic meteoroids and assume their composition is dominated by silicates we can evaluate their surface temperature at different orbital distances by equilibrating the stellar energy they absorb to the energy they emit as thermal radiation and the latent heat lost during evaporation, if they reach a high enough temperature for this process to be important \citep{Moses92}. Assuming a spherical shape we find that for radii smaller than 40 $\mu$m meteoroids reach the orbit of HD 189733 b without suffering significant evaporation, while the corresponding size for HD 209458 b is 20 $\mu$m due to the stronger flux of its star (Fig.~\ref{meteoroids}). Larger size meteoroid evaporate faster as they absorb photons more efficiently. 

This approach assumes steady state conditions which is a valid approximation for small particles as their characteristic time to reach equilibrium is small. Large size meteoroids are rare events although their frequency will depend on the age of the stellar system. For HD 209458 b and HD 189733 b with ages of 6.5$\pm$2.7 Gyr and 4.3$\pm$2.8 Gyr, respectively \citep{Boyajian14}, the meteoroid size distribution should be probably dominated by small size particles ($\leq$1mm). Thus, it is likely that at the orbit of EGPs a significant fraction of meteoroids will ablate providing a local source of silicate components such as Si and Mg, while gravitational focusing could allow for a local increase of this silicate cloud in the vicinity of the planet. However, the importance of this effect in the interpretation of the observed transit signatures of such elements \citep{Linsky10,Vidal13} depends on the meteoroid mass flux for which there is limited information.

We can further evaluate the impact such meteoroids would have once they enter the atmosphere of HD 189733 b. For this evaluation we need to know the velocity of the meteoroids relative to the planet, and the incidence angle at which they enter the atmosphere. An upper limit for the orbital velocity of the meteoroids would be their escape velocity from the star's gravity at the orbit of the planet, which for HD 189733 b is $\sim$200 km s$^{-1}$. The actual relative velocity could range between zero and this limit and in our calculations we performed estimates for relative velocities of 30 and 100 km s$^{-1}$. For the incidence angle we assume that the meteoroids enter at 45$^{\circ}$. In reality meteoroids will have a distribution over size, relative velocity, and incidence angle, but for the simple test we wish to perform here singular values will suffice. Moreover, current understanding for the meteoroid distribution properties in extrasolar systems is rather limited, and using estimates based on solar system parameters can be highly uncertain given that such distributions will depend on the whole planetary system distribution and its age. 

We calculate the ablation rate of different meteoroids by evaluating the variation of the temperature, velocity and size once they enter the atmosphere, following the approach outlined in \cite{Moses92}. Meteoroid temperature quickly increases due to collisions with the atmospheric molecules and once they reach the evaporation temperature, they start to loose mass. Our results demonstrate that at high relative velocity impacts of the order of 100 km s$^{-1}$ the meteoroid particle temperature increases faster than for the lower relative velocity (30 km s$^{-1}$). As a result, ablation for the former case occurs at pressures lower that 0.1 $\mu$bar, while for the latter particles evaporate slower and they deposit their mass near the 1 $\mu$bar pressure region (Fig.~\ref{meteoroids}). For the same relative velocity, larger size meteoroids provide broader ablation profiles because their temperature increases faster due to their larger cross section. 

The fate of the ablated material will strongly depend on the background atmospheric conditions and the abundance of the incoming meteoroids. If the temperature is low enough to allow the re-condensation of the ablated mass, silicate particles could form and affect the local extinction of the atmosphere. Apart from re-condensation, if particle's velocity is thermalised before complete ablation (as we find for smaller size meteoroids of the order of 1$\mu$m), these could directly contribute to the atmospheric opacity. Therefore there are two critical parameters that control the importance of external sources, the incoming flux of meteoroids, and the thermal structure of the upper atmosphere. Unfortunately, both parameters are not accurately constrained currently.

However the strong absorption near 10 $\mu$m due to the Si-O bond in silicate materials makes the absorption cross section of such particle to be as strong as the cross section at visible wavelengths (see Fig.~\ref{Soot_ri}). Thus, the resulting transit depth from such particles near 10 $\mu$m would be similar to the transit depth in the visible. Such a characteristic though is not supported by the observations, thus limiting the contribution of silicates for their explanation. Other metals could exist in the ablated material such as Ti, Al, Fe which also form bonds with oxygen and result in a strong absorption feature near 10 $\mu$m comparable with the absorption at visible. Therefore we can conclude that meteoroid ablation is probably not responsible for the observed transit depth of HD 189733 b. It could be important though for other exoplanets depending on the influx and atmospheric conditions. Moreover, even if not evident in the observations, meteoroid ablation could have other implications for the atmospheric properties, such as the local ionization and cloud formation. As molecules collide with high velocity meteoroids they can be ionized, while the residual or recondensed meteoroid mass could act as a nucleation site for the formation of clouds deeper in the atmosphere. 

\section{Discussion}

% Soot  Chemistry and C/O
Our calculations demonstrate that soot-composition aerosols can match the primary transit observations of HD 189733 b's upper atmosphere, while the chemical composition results suggest that the photolysis mass fluxes from the major CH$_4$ products are large enough to support the formation of soot particles. However, are the intermediate steps accessible at the conditions found in EGPs? Studies of soot formation in combustion experiments demonstrate that both aromatic and aliphatic structures are involved in the formation of nascent soot particles, while multiple mechanisms for the formation of such structures at high temperature conditions have been suggested \citep{Wang11}. Recent laboratory investigations demonstrate that CO can also increase the efficiency of aerosol production in experiments of various energy inputs \citep{Horst14}, although the mechanisms behind this interaction needs to be further investigated. Thus, the high abundance of CO in EGPs could further contribute {\color{\clr}to} the photochemical aerosol formation.

\begin{figure*}[!t]
\centering
\includegraphics[scale=0.5]{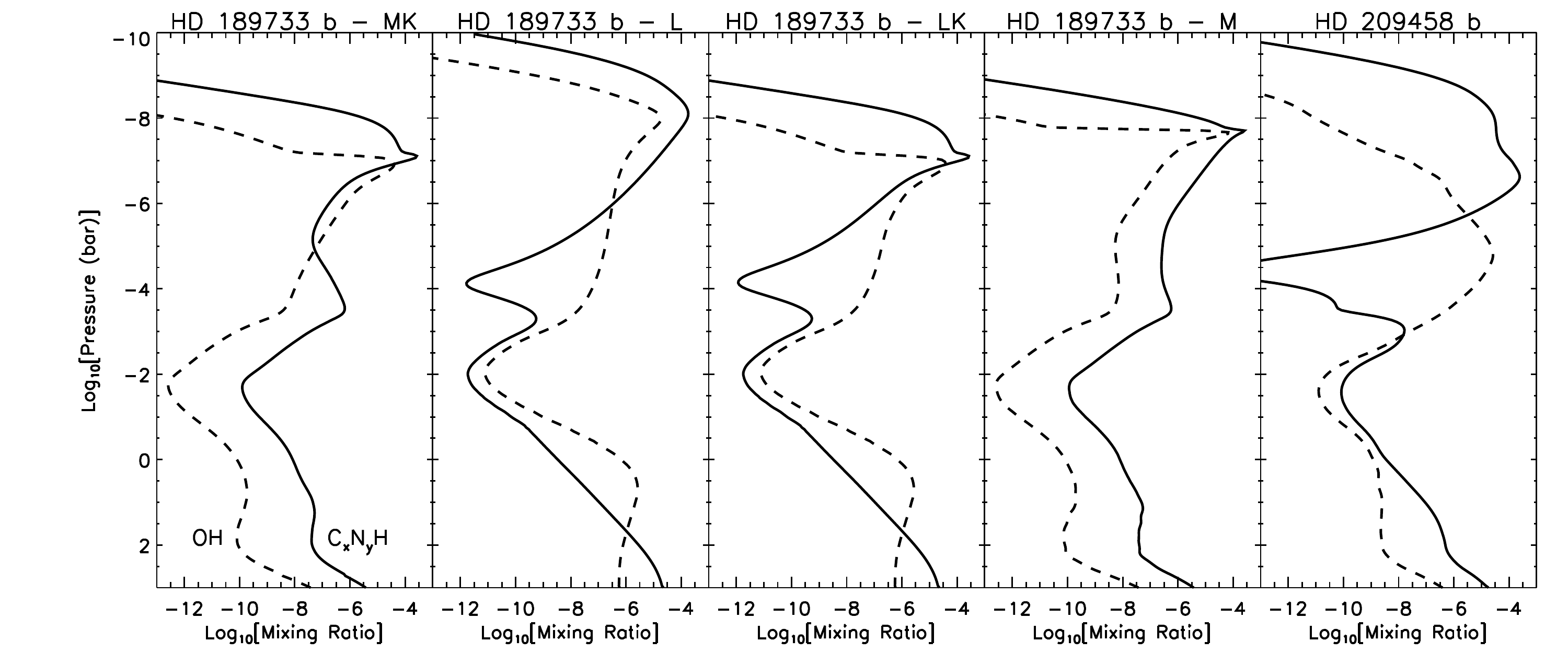}
\caption{Comparison of OH (dashed) and carbon based radical (solid) mixing ratios for the different p-T profiles assumed in the atmospheres of HD 189733 b and HD 209458 b.}\label{oxidation}
\end{figure*}

A limiting factor for the aerosol formation could be the oxidation of the atmosphere that will be controlled dominantly by the OH radicals produced in the dissociation of H$_2$O. Studies at combustion conditions demonstrate that addition of water vapor reduces the soot formation efficiency because the OH reaction with hydrocarbons limits their growth. A comparison of the OH radical abundance to that of the C/N composition radical species responsible for the growth of hydrocarbons (an approach similar to \cite{Morley13} for the evaluation of the oxidation impact), demonstrates that the latter can dominate over the former depending on the thermal structure and the atmospheric pressure (Fig.~\ref{oxidation}). This preliminary test indicates that soot formation could be possible in hot exoplanets, but a more thorough investigation considering the reactivity of each species and the impact of the radiation field is required. {\color{\clr}Nevertheless, our conclusion is in agreement with previous studies for the chemical composition of EGPs that demonstrate that reduced carbon chemistry is possible for C/O $<$ 1, at high K$_{\rm ZZ}$ values \citep{Moses13,Zahnle14}}. Moreover, studies of aerosol formation in Titan's upper atmosphere demonstrate that ion-neutral processes have a dominant role in the formation of the nascent aerosols found there, and the role of similar mechanisms on exoplanet ionospheres needs to be evaluated. Such a study goes beyond the goals of the current preliminary investigation and will be addressed in a future work. 

% Secondary eclipse
Although our results demonstrate that a photochemical aerosol distribution of soot composition can reproduce the primary eclipse observations for HD 189733 b, this result is not contradicting the anticipated clouds of silicate composition in the deeper atmosphere. The latter are still likely to form in the deeper atmosphere, but are not necessarily responsible for the heterogeneous opacity observed in the primary eclipse observations. However, the photochemical aerosols formed in the upper atmosphere of HD 189733 b could have an important role in the resulting properties of silicate clouds in the lower atmosphere, as they can act as nucleation sites. Current models for the formation of silicate composition clouds consider condensed TiO$_2$ as nucleation centers \citep{Lee17}, the latter formed through homogeneous nucleation of the gas phase. Heterogeneous nucleation is usually orders of magnitude faster than homogeneous nucleation, therefore it could significantly modify the resulting cloud particle properties. Moreover, homogeneous nucleation depends strongly on the thermal structure and it would be eliminated if the atmosphere includes the thermospheric temperature increase suggested by the atmospheric escape models. Thus, the impact of heterogeneous nucleation on the silicate condensation could also modify the vertical extend of the formed clouds. Therefore the impact of photochemical aerosols, as well as, that of the possible meteoroid ablation products, should be considered in future cloud nucleation simulations. 

% small particles or large molecules?
Current observational constraints as well as the accuracy of theoretical studies do not allow a retrieval of the thermal structure and turbulent mixing in the atmospheres of EGPs. Thus, among the different scenarios we investigated we can find solutions that provide fits to the observations. However we can note that the best fits in terms of the reduced $\chi^2$ (Table~\ref{results}) suggest that the nominal values of the eddy mixing profile provide aerosol distributions that fit the observations better. Similarly the hotter thermal structure derived from the inversion of secondary eclipse observations (L profile) also improve the fit to the observations relative to the GCM temperature profile (M profile), although the former is based on an ad hoc extension of the derived thermal structure at pressures away from the range probed in the secondary eclipse. These characteristics have further ramifications on our understanding of the shape of the particles, their charging, as well as their interaction with the atmosphere of HD 189733 b.

% fractals?
With regard to the particle shape, we found earlier that the nominal eddy profile results in smaller aerosol particle sizes with average particle radii ranging between 1.5 and 2 nm for the pressure range probed in the UV part of the spectrum (see Table~\ref{results}). As discussed above, this weak growth of the aerosol particles is driven by the strong atmospheric mixing that efficiently transports the particles to deeper pressures thereby limiting their collisions. At these small radii, particle are near their embryonic size as this is defined by the chemical mechanisms leading to their formation. Thus, their size will still be affected by the gas phase chemistry as is anticipated in both Titan's upper atmosphere \citep{Lavvas11a} and observed in soot formation laboratory experiments \citep{Wang11}. Thus, the effectively weak growth through coagulation for the nominal eddy profile and the small particle size indicate that the photochemical aerosols in the upper atmosphere of HD 189733 b are most likely spherical rather than aggregates. However, in the lower atmosphere where collisions are more frequent particles could form aggregates. The same applies for cloud particles if the mass growth rate through their collisions dominates over their growth through condensation.

So far we assumed that the particle growth is not affected by charge effects. This assumption is justified for the small particle sizes that we calculate for the nominal K$_{\rm ZZ}$ case because particles of a few nm radius will be able to acquire a limited number of charges \citep{Lavvas13}. On the other hand, the small size of the particle could potentially allow them to be drifted by electric fields in the atmosphere. A complete investigation of the problem requires a coupled investigation between aerosols, clouds and atmospheric back-ground, which is beyond the goal of our current study.

%the growth of the particles is still affected by the gas phase mechanisms leading to their formation, provided that the gaseous components partaking to this processes are still present. In other words, the mass growth through coagulation is comparable to the growth through the gas phase interaction, leading to the rounding of the formed particles. 

% Temperature
The higher temperature of the L profile makes the atmospheric scale height larger therefore the slant opacity larger relative to the M profile case. As a results the former profile provides a better fit to the observations than the latter. The discrepancy between the theoretical (M) and retrieved (L) temperature structures, suggests that a heating mechanism is required to bring the two approaches to better agreement. Our calculations demonstrate that heating by photo-absorption from soots is significant and could lead to hotter temperatures than predicted by current models. We further note that irrespective of its composition the heterogeneous opacity indicated by the primary transit observations is likely to affect both the local thermal structure and the radiation field, and thus the atmospheric photochemistry. 

Current escape calculations usually set their high pressure boundary near the 1$\mu$bar pressure level or at lower pressures, and evaluate the thermal structure of the upper atmosphere considering mainly H$_2$ and other atomic contributions, but not the impact of HCN, C$_2$H$_2$, and other strong molecular coolants. However, our model results demonstrate that the abundances of such molecules are important in the middle and upper atmosphere and can potentially affect the local temperature and the associated escape rates. Our simulated composition results do not consider the impact of the bulk escape on the mixing ratio profiles, that could affect the distribution of species in the upper atmosphere. Therefore the abundances of some species could be even higher than predicted, if the temperature conditions allow their survival. However, since our goal here is only to evaluate the production rate of photochemical aerosols that mainly depends on the column photolysis rate of different molecules, we consider that the actual shape of the species profiles in the upper atmosphere will not significantly modify our estimates. Nevertheless, the impact of molecular abundances on the thermal structure needs to be evaluated in future investigations, taking also into consideration the non-LTE character of the upper atmosphere. This aspect is not considered in the current  thermal structure models and its importance will be enhanced if molecular structures survive to significant abundances. 

We saw that soot particles vaporise at significantly higher temperatures relative to other compounds (e.g. of sulfur composition), which makes them a suitable candidate for the high temperature conditions found in HD 189733 b's atmosphere. The impact of the particle decomposition affects a part of the atmosphere that is not probed in the primary transit observations, depending on the reference pressure level of the atmosphere. However, the particle decomposition can have significant ramifications for the interaction of these particles with the background atmosphere; one example is the role of aerosols in cloud formation discussed above; another are the heterogeneous chemical processes on their surface that could have an impact on the gas phase composition. These are aspects that should be investigated in future studies and their impact evaluated against secondary eclipse observations.

\section{Conclusions}

We presented a detailed study of the possible properties of photochemical aerosols in the atmospheres of EGPs taking into account internal and external sources that could affects their production. Our results demonstrate that:
\begin{itemize}
\item	The thermal structure in the middle atmosphere affects significantly the chemical composition and the survival of molecular structures that could lead to the formation of photochemical aerosols. Moreover, the temperature profile controls the stability of possible aerosols against thermal decomposition.  
\item Among the possible candidates we considered, soot has the highest resistance to high temperatures and is a possible product in hot Jupiters  
based on the aerosol mass production rates estimated from the photochemical model results for HD 189733 b. Hotter temperatures in the upper atmosphere of HD 209458 b can explain the apparent lack of photochemical aerosols in its transmission spectrum.
\item Aerosol mass fluxes of the order of 10$^{-12}$ g cm$^{-2}$s$^{-1}$ are required to match the primary transit observations of HD 189733 b, assuming the nominal eddy mixing profile, while for lower mixing efficiency the required mass flux is reduced as particles grow to larger sizes that results in larger extinction cross sections. 
\item The average particle size for the photochemical aerosols depends strongly on the efficiency of atmospheric mixing assumed. For the nominal eddy profiles assumed in the literature according to GCMs, transport is more efficient than collisions and the resulting particle size is small between 1-2 nm in the region of the atmosphere probed by the transit observations. For these conditions particles will most likely be spherical, and their charging will not significantly affect their size distribution. If however, atmospheric mixing is weaker, particles can grow to large sizes reaching up to 100 nm for the highest mass fluxes considered. In this case aggregation is possible and charge effects need to be evaluated. 
\item The soot aerosols that match the primary transit observations would provide a significant opacity over the stellar spectrum and contribute significantly to atmospheric heating. However, the overall contribution of such particles to the thermal structure needs to be evaluated through detailed models. Moreover, soot aerosols also affect the transfer of radiation at short wavelengths and thus the atmospheric photochemistry, leading to changes in the gaseous abundances. This effect will be more pronounced at the terminators where the aerosol slant optical depth is higher. 
\item Meteoroid ablation provides a mass input in the upper atmosphere of EGPs, but the spectral signature of the heterogeneous compounds, primarily silicates, that could form is not consistent with the observations {\color{\clr}of HD 189733 b}. Thus, the $in~situ$ atmospheric composition should control aerosol formation. 
\end{itemize}


%% Included in this acknowledgments section are examples of the
%% AASTeX hypertext markup commands. Use \url without the optional [HREF]
%% argument when you want to print the url directly in the text. Otherwise,
%% use either \url or \anchor, with the HREF as the first argument and the
%% text to be printed in the second.

\acknowledgments

{\color{\clr} We thank R.V. Yelle for his comments on the manuscript. P.L. acknowledges financial support from the Programme National de Plan\'etologie (PNP-INSU) under project AMG, and from the Projet International de Coop\'eration Scientifique (PICS/CNRS) under project TAC.}

\bibliographystyle{apalike}
%\bibliography{refs}


%% in parentheses, after the keyword, but they will not be verified.



\end{document}